\documentclass[aps,prx,twocolumn,floatfix,superscriptaddress]{revtex4-2}
\usepackage[T1]{fontenc}
\usepackage[colorlinks,linkcolor=blue,citecolor=blue,urlcolor=blue]{hyperref}
\usepackage{amsmath,mleftright,mathtools,bm,amssymb}
\usepackage{dsfont}
\def\a{\alpha}
\def\b{\beta}
\def\g{\gamma}
\def\e{\epsilon}
\def\m{\mu}
\def\w{\omega}
\def\G{\Gamma}
\def\S{\Sigma}
\def\W{\Omega}
\def\nn{\nonumber}
\def\hc{\mathrm{h.c.}}
\def\Re{{\rm Re}}
\def\bS{\bar{\Sigma}}
\def\bbS{\bar{\bar{\Sigma}}}
\def\be{\bar{\epsilon}}
\def\bbe{\bar{\bar{\epsilon}}}
\def\bm{\bar{\mu}}
\def\bbm{\bar{\bar{\mu}}}
\newcommand{\cA}{\mathcal{A}}

\newcommand{\cD}{\mathcal{D}}

\newcommand{\cF}{\mathcal{F}}
\newcommand{\Ar}{\mathrm{A}}
\newcommand{\Rr}{\mathrm{R}}

\begin{document}

\title{Thermoelectric energy conversion in molecular junctions out of equilibrium}

\author{R. Tuovinen}
\affiliation{Department of Physics, Nanoscience Center, P.O. Box 35, 40014 University of Jyv{\"a}skyl{\"a}, Finland \looseness=-1}
\email{riku.m.s.tuovinen@jyu.fi}

\author{Y. Pavlyukh}
\affiliation{Institute of Theoretical Physics, Faculty of Fundamental Problems of Technology,
Wroclaw University of Science and Technology, 50-370 Wroclaw, Poland}

\begin{abstract}
Understanding time-resolved quantum transport is crucial for developing next-generation quantum technologies, particularly in nano- and molecular junctions subjected to time-dependent perturbations. Traditional steady-state approaches to quantum transport are not designed to capture the transient dynamics necessary for controlling electronic behavior at ultrafast time scales. In this work, we present a non-equilibrium Green's function formalism, within the recently-developed iterated generalized Kadanoff-Baym ansatz ($i$GKBA), to study thermoelectric quantum transport beyond the wide-band limit approximation (WBLA). We employ the Meir-Wingreen formula for both charge and energy currents and analyze the transition from Lorentzian line-width functions to the WBLA, identifying unphysical divergences in the latter. Our results highlight the importance of finite-bandwidth effects and demonstrate the efficiency of the $i$GKBA approach in modeling time-resolved thermoelectric transport, also providing benchmark comparisons against the full Kadanoff-Baym theory. We exemplify the developed theory in the calculation of time-resolved thermopower and thermoelectric energy conversion efficiency in a cyclobutadiene molecular junction.
\end{abstract}

\maketitle

\section{Introduction}

While thermoelectric devices are traditionally macroscopic and valued for their stable, time-independent performance, a new generation of applications demands precise control of charge and heat transport at atomic and molecular scales, as well as ultrafast time scales~\cite{kampfrath_resonant_2013, sengupta_terahertz_2018, gunyho_single-shot_2024}. In these regimes, conventional steady-state descriptions break down, and the interplay between energy conversion and quantum coherence must be addressed.

Recent studies have explored these challenges in a variety of systems, including ferromagnetic-superconducting interfaces~\cite{heikkila_thermal_2019}, condensed-matter platforms~\cite{pekola_colloquium_2021}, and qubit-based thermal devices~\cite{arrachea_energy_2023}. Coherence effects in mesoscopic systems enable applications such as precision thermometry and efficient refrigeration~\cite{giazotto_opportunities_2006}. Thermoelectric phenomena at the atomic and molecular scale highlight the interplay between quantum heat flow, electronic transport, and non-equilibrium thermodynamics~\cite{dubi_colloquium_2011}. Single-molecule junctions have been explored both experimentally and theoretically~\cite{reddy_thermoelectricity_2007}, with a focus on inelastic transport effects and the Seebeck coefficient~\cite{galperin_inelastic_2008a}, and the optimization of the thermoelectric figure of merit~\cite{murphy_optimal_2008, finch_giant_2009, wang_thermal_2020}. Thermoelectric efficiency and energy harvesting have been investigated in quantum-dot systems operating as thermal engines~\cite{sothmann_thermoelectric_2015, sierra_interactions_2016, yang_density_2016, mazal_nonmonotonic_2019, sobrino_thermoelectric_2023, sobrino_analytic_2025}. These advances underline the importance of thermoelectric transport in enabling quantum technologies, energy-efficient devices, and thermal control at the nanoscale.

In this work, we focus on the dynamical aspects of quantum transport, as manifested in the behavior of electrons in nano- and molecular junctions under external driving mechanisms such as time-varying voltages, electromagnetic fields, or sudden changes in system parameters~\cite{jauho_time-dependent_1994, moskalets_floquet_2002, platero_photon-assisted_2004, brandes_coherent_2005,ridley_many-body_2022}. With miniaturization to atomic and molecular scales, quantum devices reach operating frequencies in the terahertz range, corresponding to electronic time scales on the order of femtoseconds~\cite{kampfrath_resonant_2013, walowski_perspective_2016}. Optoelectronic platforms based on low-dimensional materials exploit high carrier mobility and strong light-matter coupling in this regime~\cite{viti_tailored_2021} and have demonstrated active control of electromagnetic signals on sub-picosecond scales~\cite{sengupta_terahertz_2018}. Controlling electronic behavior at ultrafast time scales~\cite{chaste_single_2008, selzer_transient_2013, zhang_ac_2014, mciver_light-induced_2020, devecchi_generation_2025} is therefore critical for designing next-generation quantum materials and technologies.

Theoretical approaches to time-dependent quantum transport can be broadly divided into two categories.
One strategy is to propagate scattering states in time~\cite{cini_time-dependent_1980,stefanucci_time-dependent_2004}, as done in microcanonical formalism~\cite{ventra_transport_2004,chien_bosonic_2012,chien_landauer_2014}, time-dependent density-functional theory~\cite{kurth_time-dependent_2005, kwok_time-dependent_2014, sobrino_thermoelectric_2021} or in density-matrix renormalization group~\cite{eckel_comparative_2010} techniques. These methods scale linearly with physical propagation time and can, in principle, capture strong correlations, but they require an explicit microscopic description of both the device and its reservoirs. This explicit treatment can become computationally expensive even for noninteracting systems~\cite{kloss_tkwant_2021} and poses conceptual challenges at finite temperature~\cite{eich_density-functional_2014,eich_functional_2017}.

The second strategy is based on non-equilibrium Green’s functions (NEGF)~\cite{danielewicz_quantum_1984, stefanucci_nonequilibrium_2025, balzer_nonequilibrium_2013}, where the dynamics are governed by the Kadanoff-Baym equations (KBE). They describe the time evolution of single-particle Green’s functions and have also found applications beyond transport, notably in ultrafast phenomena and time-resolved spectroscopy~\cite{kwong_real-time_2000, dahlen_solving_2007,myohanen_kadanoff-baym_2009, puig_von_friesen_kadanoff-baym_2010, perfetto_first-principles_2015, balzer_stopping_2016, murakami_nonequilibrium_2017, hopjan_molecular_2018, honeychurch_timescale_2019, schuler_nessi_2020}. Within this framework, transient currents can be computed using the time-dependent Meir-Wingreen~\cite{meir_landauer_1992,jauho_time-dependent_1994} or Landauer-Büttiker~\cite{landauer_spatial_1957,buttiker_generalized_1985,buttiker_four-terminal_1986,tuovinen_time-dependent_2014,ridley_photon-assisted_2025} formulas. However, direct numerical solutions of the KBE are limited by their cubic scaling with physical time, which has motivated the development of reconstruction schemes and related approximations~\cite{lipavsky_generalized_1986, kalvova_short-time_2025}. The most prominent example is the generalized Kadanoff-Baym ansatz (GKBA), which reduces the scaling to linear time~\cite{schlunzen_achieving_2020, karlsson_fast_2021, pavlyukh_time-linear_2022-1, pavlyukh_time-linear_2022, balzer_accelerating_2023, tuovinen_time-linear_2023} and performs well for charge transport in the wide-band limit~\cite{tuovinen_time-linear_2023, tuovinen_electroluminescence_2024}. However, these approximations discard essential spectral information and fail for energy currents.

Our work overcomes this bottleneck by developing the iterated GKBA ($i$GKBA), a \emph{time-linear} NEGF method that remains well defined for energy currents beyond the wide-band limit and is well suited to describe energy-dependent quantities such as heat currents. This study builds on our recent Letter~\cite{pavlyukh_open_2025}, where we employed a time-dependent Meir-Wingreen formula for energy currents and thermal transport. Here, we demonstrate that the $i$GKBA overcomes the limitations of the conventional GKBA, which fails in this regime, and provides a computationally efficient approach that retains the essential physics of the full Kadanoff-Baym equations. This enables consistent and accurate treatment of time-dependent thermoelectric transport in nanoscale junctions, capturing both transient and steady-state behavior.

\begin{figure}[t]
\center
\includegraphics[width=0.95\columnwidth]{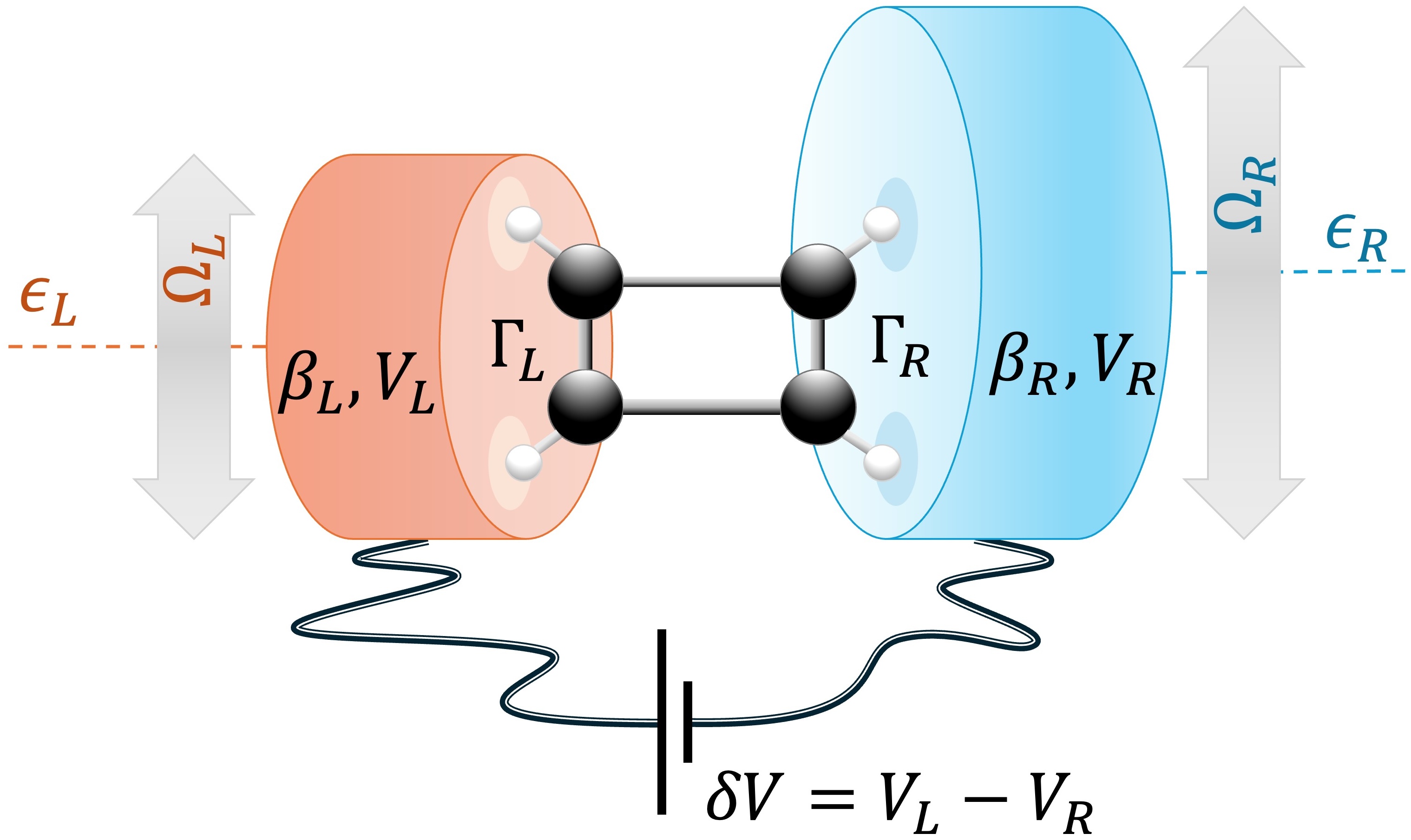}
\begin{caption}{
    Molecular junction model considered in this work. Central system (a cyclobutadiene molecule) is coupled to two leads ($\a=L,R$) via the frequency-dependent tunneling rates $\G_{\a}$, Eq.~\eqref{eq:lorentzian}, characterized by the energy centroids $\e_\a$ and bandwidths $\W_\a$. The leads are held at different temperatures $\b_\a$, and external time-dependent voltages $V_\a(t)$ are applied.
\label{fig:system}}
\end{caption}
\end{figure}

In realistic transport setups (see Fig.~\ref{fig:system}), the form of the system-lead coupling plays a decisive role. It is commonly described by a tunneling rate or line-width function, which characterizes how the states in the central region (e.g., the molecular junction) are broadened due to their interaction with the continuum of states in the leads and determines the rate at which electrons tunnel between the quantum system and the leads. The bandstructure or density of states of the leads is decisive in shaping the energy dependence of the line-width function. While the wide-band limit approximation (WBLA) is widely used for its simplicity~\cite{verzijl_applicability_2013, covito_transient_2018}, it fails to capture these effects. By concentrating on Lorentzian tunneling-rate functions, we present a detailed derivation of the iterated generalized Kadanoff-Baym ansatz and establish how it eliminates unphysical divergences in the standard GKBA, achieving benchmark agreement with full KBE solutions.

Finally, we put the developed $i$GKBA scheme to the test by investigating the real-time buildup of thermoelectric energy conversion in a molecular junction under a temperature gradient. This reveals transient efficiency windows, i.e., periods where performance can exceed the steady state, establishing dynamics as essential for understanding nanoscale energy conversion and for guiding the development of future quantum and energy-harvesting technologies.

\section{Quantum transport setup and observables}

\subsection{The Hamiltonian}
We consider a generic quantum-transport model, consisting of a central system coupled to metallic leads shown in Fig.~\ref{fig:system}, described by the time-dependent Hamiltonian
\begin{multline}\label{eq:ham}
\hat{H}(t) = \sum_{mn}h_{mn}(t)\hat{c}_m^\dagger \hat{c}_n + \frac{1}{2}\sum_{mnpq}v_{mnpq}(t)\hat{c}_m^\dagger\hat{c}^\dagger_n\hat{c}_p\hat{c}_q \\
 + \sum_{k\a} E_{k\a}(t)\hat{c}_{k\a}^\dagger\hat{c}_{k\a} + \sum_{mk\a}\left( T_{mk\a}(t)\hat{c}_m^\dagger\hat{c}_{k\a} + \hc \right) ,
\end{multline}
where $\hat{c}_m^\dagger / \hat{c}_n$ create/annihilate electrons at the central region (orbital/spin basis $m,n$), and $\hat{c}_{k\a}^\dagger / \hat{c}_{k\a}$ create/annihilate electrons at leads (momentum/spin state $k$ at lead $\a$).
Here, $h_{mn}$ and $v_{mnpq}$ are the one- and two-electron matrix elements of the central region, $E_{k\a}$ is the energy dispersion of the leads, and $T_{mk\a}$ the tunneling matrix elements between the central region and leads.

The time-dependence of Eq.~\eqref{eq:ham} describes an out-of-equilibrium transport process generated by, e.g., a voltage or coupling switch. The lead energies are modified by a time-dependent bias voltage as $E_{k\a}(t) = E_{k\a} + V_\a(t)$. Similarly, a time-dependent (possibly non-local) potential profile within the central region is modeled by $h_{mn}(t) = h_{mn}+u_{mn}(t)$. In addition, we allow for the coupling and interaction matrix elements, $T_{mk\a}(t)=T_{mk\a}s_\a(t)$ and $v_{mnpq}(t)=v_{mnpq}s_v(t)$ to also have a time-parameter to describe, e.g., adiabatic switching through some time-dependent ramp functions $s(t)$.

\subsection{Non-equilibrium Green's function theory}

The central object of the non-equilibrium Green's function theory, in describing the out-of-equilibrium quantum-transport setting of Eq.~\eqref{eq:ham} is the one-electron \emph{lesser} ($<$) Green's function (GF)~\cite{stefanucci_nonequilibrium_2025}
\begin{equation}\label{eq:gf}
G_{rs}^<(t,t') = i \left\langle \hat{c}_s^\dagger(t') \hat{c}_r(t) \right\rangle,
\end{equation}
where the labels $r,s$ refer to either the lead or central region states, the creation/annihilation operators are represented in the Heisenberg picture, and the ensemble average $\langle \cdots \rangle$ is taken as a trace over the equilibrium density matrix. This function naturally captures quantum transport processes, as electrons continuously hop onto and off the central molecule, thereby changing the particle number in the central region. The associated one-electron density matrix is obtained as the equal-time limit $\rho(t) \equiv -i G^<(t,t)$.

The GF obeys the integro-differential Kadanoff-Baym equation (in matrix form)
\begin{align}\label{eq:gf-eom}
& \left[i \partial_t - h^{\mathrm{HF}}(t) \right]G^<(t,t') \nonumber \\
& = \int d \bar{t} \left[\Sigma^<(t,\bar{t})G^\Ar(\bar{t},t') + \Sigma^\Rr(t,\bar{t})G^<(\bar{t},t')\right],
\end{align}
where $h_{mn}^{\mathrm{HF}}(t) = h_{mn}(t)+V_{mn}^{\mathrm{HF}}(t)$ is the one-electron Hamiltonian including the Hartree-Fock (HF) potential $V_{mn}^{\mathrm{HF}}(t) = \sum_{pq}(v_{mpqn}(t)-v_{mpnq}(t))\rho_{qp}(t)$. The self-energy $\Sigma=\Sigma_c+\Sigma_{\mathrm{em}}$ is composed of the correlation and embedding parts. The correlation self-energy can be specified at different levels of approximations: second-order Born~\cite{dahlen_solving_2007, balzer_electronic_2012}, $GW$ and $T$-matrix~\cite{stan_levels_2009, puig_von_friesen_kadanoff-baym_2010}, or based on the non-equilibrium dynamical mean-field theory~\cite{freericks_nonequilibrium_2006, aoki_nonequilibrium_2014, strand_nonequilibrium_2015}. The embedding self-energy accounts for the system being open to an environment consisting of the leads. The superscripts `$\Ar$' and `$\Rr$' denote the advanced and retarded components, respectively.

Extracting the equal-time limit of Eq.~\eqref{eq:gf-eom} gives the equation of motion for the density matrix~\cite{tuovinen_time-linear_2023}
\begin{equation}\label{eq:rho-eom}
i \frac{d}{d t} \rho(t) = \left[h^{\mathrm{HF}}(t)\rho(t)-i I_c(t) - i I_{\mathrm{em}}(t)\right] - \hc ,
\end{equation}
where $I_c$ and $I_{\mathrm{em}}$ are the collision integrals (convolutions) of the correlation and embedding self-energies, respectively, with the GF of the central region. Analogously, interactions of electrons with phonons and photons can be incorporated in the formalism leading to a set of coupled equations for the electron and boson correlators~\cite{sakkinen_many-body_2015, schuler_time-dependent_2016, karlsson_non-equilibrium_2020}.

The equation of motion for the density matrix in Eq.~\eqref{eq:rho-eom} is not closed because the collision integrals depend functionally on the two-time Green's functions and self-energies. In Sec.~\ref{sec:devel}, we use the \emph{reconstruction equation}~\cite{lipavsky_generalized_1986} to access the two-time lesser GF in terms of simpler propagators, which allows us to develop a time-linear scaling framework.

\subsection{Charge and energy currents}\label{sec:obs}

We define a current operator for lead $\a$ satisfying the Heisenberg equation of motion
\begin{equation}
\hat{J}_{\a}^{\nu}(t) = \frac{d \hat{H}_\a^{\nu}}{d t} = i \left[\hat{H},\hat{H}_\a^{\nu}\right] ,
\end{equation}
where $\hat{H}_\a^{\nu}\equiv \sum_{k}E_{k\a}^\nu \hat{c}_{k\a}^\dagger \hat{c}_{k\a}$ corresponds to the $\a$-lead part of the Hamiltonian~\eqref{eq:ham} when $\nu=1$ and to the particle number operator $\hat{N}_\a \equiv \sum_k \hat{c}_{k\a}^\dagger\hat{c}_{k\a}$ when $\nu=0$. The commutator has non-zero entries only with the coupling block of the full Hamiltonian~\eqref{eq:ham} and we obtain for the currents, $J_\a^{\nu}\equiv \langle \hat{J}_\a^{\nu} \rangle$,
\begin{multline}\label{eq:current-deriv}
J_\a^{\nu}(t) = \sum_{mk}E_{k\a}^\nu\left[T_{mk\a}(t)i\bigl\langle\hat{c}_m^\dagger\hat{c}_{k\a}\bigr\rangle \right.\\
\left.-T_{k\a m}^*(t)i\bigl\langle\hat{c}_{k\a}^\dagger\hat{c}_m\bigr\rangle\right],
\end{multline}
where the operators are understood in the Heisenberg picture. Here, we identify the equal-time lesser GF of the lead-molecule block, cf.~Eq.~\eqref{eq:gf}: $G_{k\a m}^<(t,t) = -\left(G_{m k\a}^<(t,t)\right)^* = i \bigl\langle \hat{c}^\dagger_m(t)\hat{c}_{k\a}(t) \bigr\rangle$. For symmetric and real tunneling matrices, we may rewrite Eq.~\eqref{eq:current-deriv} as~\cite{eich_luttinger_2014}
\begin{equation}
J_\a^{\nu}(t) = 2\sum_{mk}E_{k\a}^\nu\Re\left[T_{mk\a}(t)G_{k\a m}^<(t,t)\right].
\end{equation}
By using the equations of motion of the lead-molecule Green's function with Langreth rules we obtain the Meir-Wingreen formula for the charge and energy current (see Ref.~\cite{ridley_many-body_2022})
\begin{multline}\label{eq:meir-wingreen}
J_\a^{\nu}(t)  = 2 \Re \text{Tr} \int_{-\infty}^\infty d t' \left[\Sigma_\a^{\nu,<}(t,t')G^{\mathrm{A}}(t',t) \right.\\
\left.+\Sigma_\a^{\nu,\mathrm{R}}(t,t')G^<(t',t)\right] \equiv 2 \Re \text{Tr}I_\a^\nu(t),
\end{multline}
where the lead self-energies $\Sigma_\a^{\nu}$ will be defined in detail in Sec.~\ref{sec:devel}.
For convenience, we also denote $J_\a(t)\equiv J_\a^{0}(t)$ as the charge current and $J_\a^{\mathrm{E}}(t)\equiv J_\a^{1}(t)$ as the energy current. The associated heat current is then found as $J^H_\alpha(t) = J^E_\alpha(t) - \mu_\alpha J_\a(t)$, where $\mu_\a$ is the chemical potential of lead $\a$.

\section{Time-linear formulation with Lorentzian leads}\label{sec:devel}

\subsection{Collision integrals}

The correlation self-energy terms appearing in the collision integrals [see Eq.~\eqref{eq:rho-eom}] can be separated and dealt with independently for various many-body effects~\cite{schlunzen_achieving_2020, karlsson_fast_2021, pavlyukh_time-linear_2022-1}. Focusing now on the embedding self-energy, it accounts for all processes in which an electron transitions from orbital $m$ of the central molecule to an energy level $k$ in lead $\a$, and subsequently returns to orbital $n$ of the molecule. It is thus specified in terms of the lead and coupling Hamiltonians, by a summation over all leads, $\Sigma_{\mathrm{em}} = \sum_\a \Sigma_\a^0$ with $\Sigma_{\a,mn}^0(t,t') = \sum_k T_{mk\a}(t)g_{k\a}(t,t')T_{k\a n}(t')$, where the lead Green's function $g$ is taken to be non-interacting~\footnote{The retarded and lesser components of the lead GF satisfy the equations of motion: $\left[i\partial_t-E_{k\a}(t)\right]g^\Rr(t,t')=\delta(t-t')$, $\left[i\partial_t-E_{k\a}(t)\right]g^<(t,t')=0$.}. Similarly for the collision integral appearing in Eq.~\eqref{eq:rho-eom}, $I_{\mathrm{em}}(t)=\sum_\a I^0_\a(t)$.

Let us now consider the collision integrals defined by Eq.~\eqref{eq:meir-wingreen}
\begin{align}
  \label{eq:collision}
 I_\a^\nu(t)=\int \! dx \bigl[\Sigma_\a^{\nu,<}(t,x) G^{\Ar}(x,t)+\Sigma_\a^{\nu,\Rr}(t,x)G^<(x,t)\bigr].
\end{align}
We will utilize symbols $x,y,z$ for intermediate times. Dealing with the lesser GF component in Eq.~\eqref{eq:collision} is a fundamental problem of the NEGF formalism. Our goal is to derive a set of coupled equations of motion (EOM) for Eq.~\eqref{eq:collision}, to be co-evolved with Eq.~\eqref{eq:rho-eom}, constituting a time-linear scaling framework.

\subsection{Reconstruction equations}

It is convenient to split the lesser GF into the retarded- and advanced-like ($\Rr$/$\Ar$) components 
\begin{align}
  G^<(t,t')&=\theta(t-t')G^<_{\Rr}(t,t')-\theta(t'-t)G^<_{\Ar}(t,t'). \label{eq:split}
\end{align}
For these components, we will use of the reconstruction equations derived by Lipavsk\'{y}, \v{S}pi\v{c}ka, and Velick\'{y}~\cite{lipavsky_generalized_1986}:
\begin{multline}
  G^<_{\Rr}(t, x) = - G^{\Rr}(t, x)\rho(x)
  -\int \! d{y}\, G^{\Rr}(t, y) \theta(y-x)\\
  \times \left\{ \int\! d{z}\,\S^<(y, z)G^{\Ar}(z, x) - \S^{\Rr}(y, z)G^<_{\Ar}(z, x)\right\},
  \label{eq:reconstr:R}
\end{multline}
and 
\begin{multline}
  G^<_{\Ar}(x, t) = -\rho(x) G^{\Ar}(x,t)\\
   -\int d{y} \left\{ \int\! d{z}\,G^{\Rr}(x, z)\S^<(z, y) +G^<_{\Rr}(x, z) \S^{\Ar}(z, y)\right\}\\
   \times \theta(y-x)G^{\Ar}(y,t).
   \label{eq:reconstr:A}
\end{multline}
One obtaines the ``standard'' GKBA equations by neglecting the memory corrections (given by the integral over $y$). The iterated GKBA ($i$GKBA) is obtained by iterating the equations further, i.e., inserting the GKBA expression for $G^<_{\Ar}$ from the second equation into the first, and $G^<_{\Rr}$ from the first equation into the second. In order to efficiently deal with different kinds of convolution integrals that appear in this work, we introduce short-hand notations:
\begin{align}
  \bigl[a f\cdot b\bigr](t,t')&=\smallint\! d{x}\,a(t,x)f(x)b(x,t'),\\[9pt]
  \bigl[a f\cdot b\bigr]_{\Rr/\Ar}(t,t')&=\theta(\pm(t-t'))\bigl[a f\cdot b\bigr](t,t').
\end{align}
This allows to rewrite the $i$GKBA equations as follows:
\begin{subequations}
  \label{eq:iGKBA:reconstr}
  \begin{align}
    G^<_{\Rr}(t, x) &= - G^{\Rr}(t, x)\rho(x)\nn\\
    &\,+\Bigl[G^{\Rr}\cdot \bigl[\S^< \cdot G^{\Ar} +\S^{\Rr}\rho\cdot G^{\Ar}\bigr]_{\Rr}\Bigr](t,x),\\
    G^<_{\Ar}(x,t) &= -\rho(x) G^{\Ar}(x,t)\nn\\
    &\, -\Bigl[\bigl[ G^{\Rr}\cdot \S^< -G^{\Rr}\rho\cdot\S^{\Ar}\bigr]_{\Ar}\cdot G^{\Ar}\Bigr](x,t).
    \label{eq:G<A}
  \end{align}
\end{subequations}
Equations~\eqref{eq:iGKBA:reconstr} constitute the first important ingredient of our formalism enabling us to evaluate the embedding collision integral.

\subsection{Embedding self-energy components}

The embedding self-energy is the second important ingredient of our theory. It encodes microscopic properties of the leads (numbered by the index $\a$) and is based on the form of the coupling $T_{mk\a}$ between the state $m$ of central region to the state $k$ of the lead $\a$ with energy $E_{k\a}$. From a mathematical point of view, the Lorentzian line-width function~\cite{tang_full-counting_2014}
\begin{multline}\label{eq:lorentzian}
  \Gamma_{\a,mn}(\w) = 2\pi\sum_{k} T_{mk\a}\delta(E_{k\a}-\w)T_{k\a n}\\
  =\gamma_{\a,mn}\frac{\Omega_\a^2}{(\w-\e_\a)^2+\Omega_\a^2},
\end{multline}
offers a number of technical advantages. In the equation above, $\e_\a$ and $\Omega_\a$ are the energy centroid and bandwidth, respectively, whereas $\gamma_{\a}$ is a constant matrix describing the coupling configuration. The wide-band limit approximation is obtained when the bandwidth approaches infinity $\Omega_\alpha\to\infty$, in which case the line-width function becomes frequency independent $\Gamma_\alpha(\w)\approx \gamma_\a$. Since the tunneling matrix elements acquire time-dependence $T_{mk\a} s_\a(t)$ via the ramp functions $s_\a(t)$ for lead $\a$, a time-dependent prefactor appears
\begin{align*}
  s_\a(t) e^{-i\phi_\a(t,t')}s_\a(t')& = s_\a(t) u_\a(t,t') s_\a(t'),
\end{align*}
where $\phi_\a(t,t')\equiv \int_{t'}^t d{x} V_\a(x)$ is the accumulated phase due to the applied voltage. With these ingredients, different components of the embedding self-energy can be constructed as Fourier transforms, $\cF[a](\tau)=\int\frac{d \w}{2\pi}e^{-i\w\tau}a(\w)$, of the lead tunneling rates~\cite{stefanucci_nonequilibrium_2025}:
\begin{subequations}
  \label{eq:sgm:emb}
\begin{align}
  \Sigma_\a^{\nu, \Rr}(t, t') & = \left[\Sigma_\a^{\nu, \Ar}(t', t) \right]^\dagger =-i s_\a(t)u_\a(t,t') s_\a(t') \nn\\
   & \qquad \times \cF\mleft[\w^\nu \Gamma_\a(\w)\mright](t-t')\theta(t-t'),\label{eq:sigmar}\\
  \Sigma_\a^{\nu, <}(t,t') & = is_\a(t)u_\a(t,t') s_\a(t')\nn\\
  & \qquad \times \cF\mleft[\w^\nu f_\a(\w)\Gamma_\a (\w)\mright](t-t'),\label{eq:sigmalss}
\end{align}
\end{subequations}
where $f_\a(\w)$ is the Fermi-Dirac distribution function
\begin{align}
  f_\a(\w)&=\frac{1}{e^{\beta_\a(\w-\mu_\a)}+1},
  \end{align}
for inverse temperature $\beta_\a$ and chemical potential $\mu_\a$. 

We will first address the embedding self-energy in the calculation of charge currents and lighten the notation $\Sigma_\a^{0}=\S_\a$. Due to the relatively simple form of the line-width function~\eqref{eq:lorentzian}, by closing the integration contour in the complex lower, upper half-plane we obtain for $\S_\a^{\Rr}(t,x)= s_\a(t)\bS_\a^{\Rr}(t,x)$,  $\S_\b^{\Ar}(x, t)=\bS_\b^{\Ar}(x, t)s_\b(t)$, respectively:
\begin{subequations}
\begin{align}
  \bS_\a^{\Rr}(t, x)&= -\frac{i}{2} \W_\a\gamma_\a u_\a(t, x)s_\a(x)e^{-i\be_\a (t-x)}\theta(t-x),\label{eq:S:R:fin}\\
  \bS_\b^{\Ar}(x, t)&=\frac{i}{2} \W_\b \gamma_\b e^{i\be_\b^*(t-x)}s_\b(x)u_\b(x, t) \theta(t-x),\label{eq:S:A:fin}
\end{align}
\end{subequations}
where we introduced $\be_\a=\e_\a-i\W_\a$, and $1\le \a,\b\le N_{\text{leads}}$. By explicit calculation, we find the EOMs:
\begin{subequations}
  \label{eom:SRA}
  \begin{align}
   i\frac{d}{dt}\bS_\a^{\Rr}(t, x)&=\bbe_\a(t)\bS_\a^{\Rr}(t, x)+\frac{\W_\a}{2}\g_\a s_\a(t)\delta(t-x),\\
    -i\frac{d}{dt}\bS_\b^{\Ar}(x, t)&=\bbe_\b^*(t)\bS_\b^{\Ar}(x,t)+\frac{\W_\b}{2}\g_\b s_\b(t)\delta(x-t),
\end{align}
\end{subequations}
where $\bbe_\a(t)=V_\a(t)+\be_\a$, and the equal-time conditions read $\bS_\a^{\Rr}(t, t)= \left[\bS_\a^{\Ar}(t, t)\right]^\dagger= -\frac{i}{4} \W_\a\gamma_\a s_\a(t)$.

In order to compute the lesser self-energy component we use a standard approach of representing the fermionic distribution function in form of a pole expansion~\cite{hu_communication:_2010}
\begin{multline}
  f_\a(\omega)=\frac12-\sum_{\ell\ge 1}^{N_p}\eta_\ell
  \Bigl[\frac{1}{\beta_\a(\w-\mu_\a)+i\zeta_\ell}\\+\frac{1}{\beta_\a(\w-\mu_\a)-i\zeta_\ell}\Bigr],\quad\text{with\; $\zeta_\ell>0$.}
  \label{eq:pole:exp}
\end{multline}
Introducing $\S_{\Rr,\a}^{<}(t,x)= s_\a(t)\bS_{\Rr,\a}^{<}(t,x)$,  we obtain after contour integrations the explicit expression:
\begin{align}
  \bS_{\Rr,\a}^{<}(t, x)&
=u_\a(t,x) s_\a(x)\sum_{\ell\ge0}\bar{\eta}_{\a\ell}e^{-i\bar{\mu}_{\a\ell}(t-x)}\nn\\
&\qquad=\sum_{\ell\ge0}\bar{\eta}_{\a\ell}\bbS_{\Rr,\a\ell}^{<}(t, x), \label{eq:S:<:fin}
\end{align}
where the expansion coefficients 
\begin{align}
  \bar{\eta}_{\a\ell}&=\begin{cases}
  i\frac{\gamma_\a}{2}\W_\a f_\a(\e_\a-i\W_\a)& \ell=0,\\
  -\frac{\eta_\ell}{\b_\a}\G_\a\mleft(\m_\a-i\frac{\zeta_\ell}{\b_\a}\mright)&\ell\ge 1;
  \end{cases}
\end{align}
and the exponents $\bar{\mu}_{\a\ell}$ are given by
\begin{align}
  \bar{\mu}_{\a\ell}&=
  \begin{cases}
  \e_\a-i\W_\a& \ell=0,\\
  \m_\a-i\frac{\zeta_\ell}{\b_\a}&\ell\ge 1.
  \end{cases}
\end{align}
For the lesser self-energy of advanced type we obtain:
\begin{align}
  \bS_{\Ar,\b}^{<}(x, t)&=u_\b(x, t) s_\b(x)\sum_{\ell\ge0}\bar{\eta}^*_{\b\ell}e^{i\bar{\mu}^*_{\b\ell}(t-x)}\nn\\
&\qquad=\sum_{\ell\ge0} \bar{\eta}^*_{\b\ell}\bbS_{\Ar,\b\ell}^{<}(x, t) .
\end{align}
The partial self-energies fulfill the EOMs ($\bbm_{\a\ell}(t)=V_\a(t)+\bm_{\a\ell}$):
\begin{subequations}
  \label{eom:S<}
\begin{align}
  i\frac{d}{dt}\bbS_{\Rr,\a\ell}^{<}(t, x)&=\bbm_{\a\ell}\bbS_{\Rr,\a\ell}^{<}(t, x),\\
  -i\frac{d}{dt}\bbS_{\Ar,\b\ell}^{<}(x, t)&=\bbm_{\b\ell}^*\bbS_{\Ar,\b\ell}^{<}(x, t),
\end{align}
\end{subequations}
and the equal-time condition reads $\bbS_{\Rr,\a\ell}^{<}(t, t)=\bbS_{\Ar,\a\ell}^{<}(t, t)=s_\a(t)$.

So far, we dealt with ordinary self-energies, i.\,e., the $\nu=0$ case. Starting from the Fourier representations~\eqref{eq:sgm:emb}, the relations between $\nu=0$ and $\nu=1$ self-energies can be established:
\begin{align}
  \bS_\a^{1,\Rr}(t,x)&=\be_\a\bS_\a^{\Rr}(t,x),\\
  \bS_{\Rr,\a}^{1,<}(t,x)&=\sum_{\ell\ge0}\bm_{\a\ell}\bar{\eta}_{\a\ell}\bbS_{\Rr,\a\ell}^{<}(t,x).
\end{align}

\subsection{Working scheme}

With all the ingredients, the two cases ($\nu=0,1$) of the collision integral~\eqref{eq:collision} can be written together as:
\begin{multline}
  I^{\nu}_\a(t)=  s_\a(t)\Biggl\{\sum_{\ell\ge0} \bm_{\a\ell}^\nu\bar{\eta}_{\a\ell} \cD_{\a\ell}^{c}(t) +\be_\a^\nu\cD_\a^{d}(t)\\
  +\be_\a^\nu\sum_\b \bigl[\cA_{\a\b}^a-\cA_{\a\b}^b\bigr](t)\Biggr\},\label{eq:I:nu}
\end{multline}
where new correlators have been introduced: The first term in the collision integral~\eqref{eq:collision} gives rise to
\begin{align}
  \cD_{\a\ell}^{c}(t)&=\bigl[\bbS_{\Rr,\a\ell}^<\cdot G^{\Ar}\bigr](t,t).  \label{def:Dc}
\end{align}
The second collision term in combination with the first term of the reconstruction equation~\eqref{eq:G<A} leads to
\begin{align}
  \cD_{\a}^{d}(t)&=\mleft[ \bS^{\Rr}_{\a}\rho\cdot G^{\Ar}\mright](t,t). \label{def:Dd}
\end{align}
Finally, the correction terms in Eq.~\eqref{eq:iGKBA:reconstr} give rise to 
\begin{subequations}
  \label{def:A}
\begin{align}
  \cA_{\a\b}^{a}(t)&=\mleft[\bS_\a^{\Rr}\cdot\big[G^{\Rr}\cdot\S^<_\b\big]_{\Ar}\cdot G^{\Ar}\mright](t, t),\\
  \cA_{\a\b}^{b}(t)&=\mleft[\bS_\a^{\Rr}\cdot \big[G^{\Rr}\rho\cdot\S^{\Ar}_\b\big]_{\Ar}\cdot G^{\Ar}\mright](t,t).
  \label{def:Aa}
\end{align}
\end{subequations}

\begin{figure}
 \includegraphics[width=0.95\columnwidth]{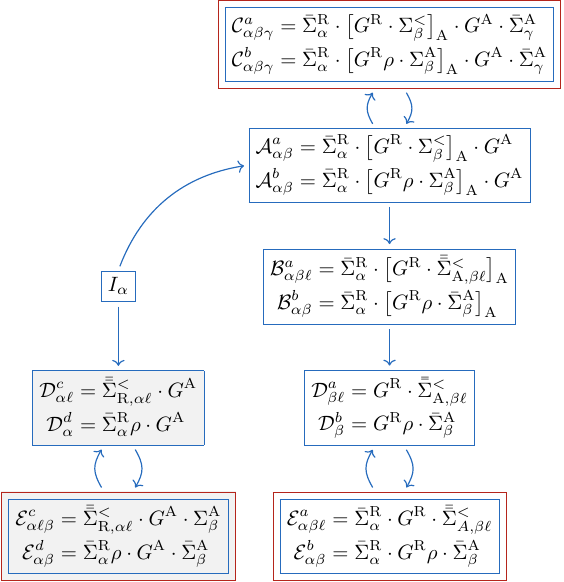}
\begin{caption}{
      Derivation of the EOMs for the $i$GKBA correlators. Correlators that originate from the self-energy terms in Eqs.~\eqref{eq:eom:GRA} are framed in red. Correlators that appear already at the GKBA level are shaded. \label{fig:diagram}}
\end{caption}
\end{figure}

Complementing the EOMs for different self-energy components with the EOMs for retarded 
\begin{subequations}
  \label{eq:eom:GRA}
\begin{multline}
  i\frac{d}{dt} G^{\Rr}(t, t')= \delta(t-t')\\ + h^{\mathrm{HF}}(t) G^{\Rr}(t, t')
  +\bigl[\S^{\Rr}\cdot G^{\Rr}\bigr](t, t'),\label{eq:eom:GR}
\end{multline}
and advanced electronic propagators
\begin{multline}
  -i\frac{d}{dt} G^{\Ar}(t', t)=\delta(t'-t)\\+G^{\Ar}(t', t)h^{\mathrm{HF}}(t)
  +\bigl[G^{\Ar}\cdot \S^{\Ar}\bigr](t',t),\label{eq:eom:GA}
\end{multline}
\end{subequations}
one can formulate a system of EOMs for the correlators, Fig.~\ref{fig:diagram}. As explained in Ref.~\cite{pavlyukh_open_2025}, this requires intermediate correlators $\mathcal{B}, \mathcal{C}, \mathcal{E}$. We refer to the original work for the full set of 14 coupled equations. It is important to note, however, that these equations simplify considerably if the last convolution terms in Eqs.~\eqref{eq:eom:GRA} are neglected. In this case, correlators in red frames are not considered.

\section{Benchmark simulations for a quantum-dot system}

There are different levels of approximations outlined in Sec.~\ref{sec:devel}. Inserting the GKBA iteratively back into the reconstruction equations~\eqref{eq:reconstr:R} and~\eqref{eq:reconstr:A} resulted in the full set of $14$ EOMs, which is referred to $i$GKBA. Neglecting the correction terms $\mathcal{A}$ in the collision integral~\eqref{eq:I:nu} reduces to the GKBA framework (only shaded correlators in Fig.~\ref{fig:diagram} are considered) with the embedding self-energies described beyond the WBLA, through the Lorentzian line-width function~\eqref{eq:lorentzian}. This finite-bandwidth Lorentzian description can be applied systematically to both the collision integrals~\eqref{eq:I:nu} and the equations of motion for the advanced propagators~\eqref{eq:eom:GRA}. We refer to this scheme as ``GKBA ($\Omega, \ G^{\mathrm{A}}$)''. Alternatively, the retarded/advanced propagators can be approximated at the wide-band approximation level, i.e., the bandwidth $\Omega$ is finite for the embedding self-energies, but the equations of motion~\eqref{eq:eom:GRA} include an effective Hamiltonian $h_{\mathrm{eff}}=h^{\mathrm{HF}}-i \gamma/2$ with the frequency-independent line-width matrix ($\gamma=\sum_\a\gamma_\a$; see Eq.~\eqref{eq:lorentzian}). We refer to this as ``GKBA ($\Omega, \ h_{\mathrm{eff}}$)'' with red-framed correlators in Fig.~\ref{fig:diagram} being additionally neglected. The limiting case, $\Omega\to\infty$ is the wide-band limit approximation, simply referred to as ``GKBA (WBLA)'': For the charge density and current, we refer to Ref.~\cite{tuovinen_time-linear_2023} and for the energy current, an alternative derivation is presented in the Supplemental Material~\cite{suppmat}. The ($i$)GKBA calculations were performed using the \textsc{cheers} code~\cite{pavlyukh_cheers_2024, zenodo}.

We focus on a thermoelectric transport setup consisting of a non-interacting two-terminal quantum dot with a temperature gradient and a time-dependent chemical potential difference. This model, studied analytically in Ref.~\cite{kara-slimane_simulating_2020}, can capture complex dynamic thermoelectric behavior, and our benchmark calculations illustrate how different operating regimes impact efficiency, charge and energy-current responses of the system, and the validity range of the developed theoretical formulation in Sec.~\ref{sec:devel}.

\begin{figure}[t]
\center
\includegraphics[width=0.95\columnwidth]{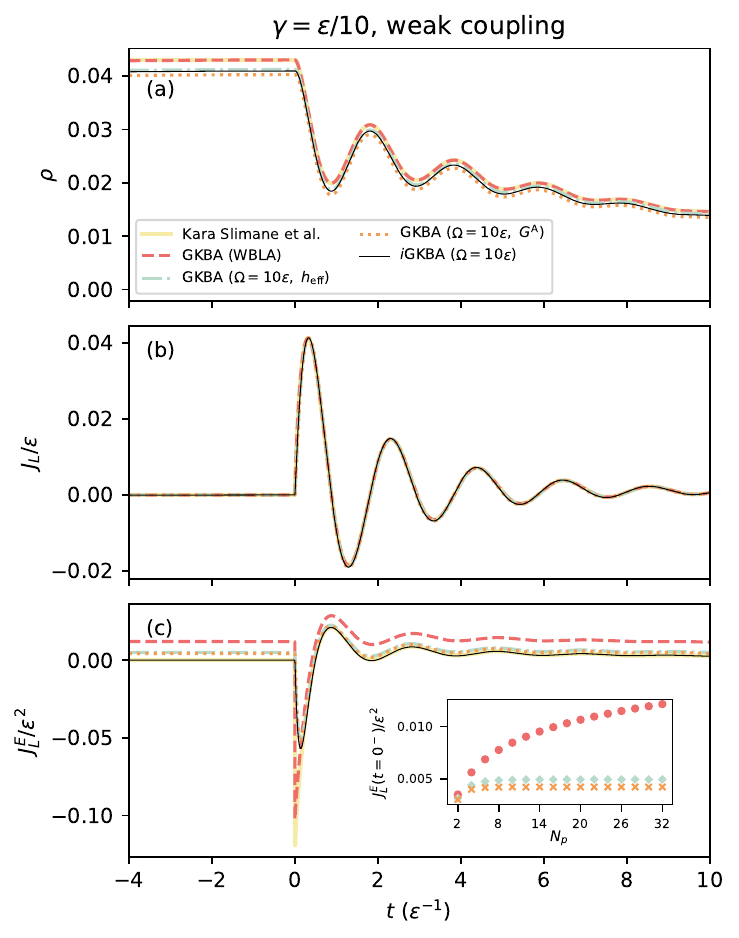}
\begin{caption}{
Time-dependent charge density $\rho$ (a), charge current $J_L$ (b), and energy current $J_L^E$ (c) at the left-lead interface with the quantum dot being weakly coupled to the leads, $\gamma=\varepsilon/10$, and driven by a sudden gate voltage $u(t)=(5\varepsilon/2)\theta(t)$. Inset in panel (c) displays the equilibrium energy current, before the sudden gate voltage is switched on, in terms of the number of poles. Other parameters are $\mu_L=\mu_R=0$, $\beta_L=\beta_R=10/\varepsilon$.
\label{fig:weak}}
\end{caption}
\end{figure}

Referring back to Eq.~\eqref{eq:ham}, we set the quantum dot energy level to $h(t)=\varepsilon/2+u(t)$, where a sudden gate voltage is applied as $u(t)=(5\varepsilon/2)\theta(t)$. All energies are expressed in units of $\varepsilon>0$, defined relative to the average chemical potential. The quantum dot is non-interacting, i.e., $v=0$ in Eq.~\eqref{eq:ham}. The chemical potentials of the left and right leads are first set to $\mu_L=\mu_R=0$ and there is no temperature gradient $\beta_L=\beta_R=10/\varepsilon$. In all calculations, we set $\Omega_\a=\Omega$ for all leads $\a$. In Fig.~\ref{fig:weak}, the time-dependent response is analyzed in terms of the charge density on the quantum dot $\rho$ and the charge and energy current at the left-lead interface $J_L$, $J_L^E$, when the quantum dot is weakly coupled to the leads $\gamma_L+\gamma_R=\gamma=\varepsilon/10$. We note that $\gamma$ is a scalar for the single-level quantum-dot system. The time evolutions are started from an initially disconnected system $\rho(t\to -\infty)=0$ and adiabatically ramping the coupling between the quantum dot and the leads. As discussed in Ref.~\cite{pavlyukh_open_2025}, with aligned chemical potentials and no temperature gradients, the equilibrium state should exhibit no currents. The comparison of density and charge current to the analytical result of Ref.~\cite{kara-slimane_simulating_2020} (``Kara Slimane et al.'' in Fig.~\ref{fig:weak}) is excellent already at the WBLA level. When the leads are described with a finite-width Lorentzian, GKBA ($\Omega=10\varepsilon, \ h_{\mathrm{eff}}$) and GKBA ($\Omega=10\varepsilon, \ G^{\mathrm{A}}$), there is a small change in the equilibrium density $\rho(t=0)$. This has no apparent effect on the charge current. The $i$GKBA results follow the same behavior.

For the energy current, Fig.~\ref{fig:weak}(c), the situation changes qualitatively. Although the result of Ref.~\cite{kara-slimane_simulating_2020} is obtained by numerical integration, the energy current still converges to a physically expected value for a two-lead setup at equilibrium $J_L^E(t=0)=0$. The $i$GKBA calculation matches this fairly well, although the finite-bandwidth $\Omega=10\varepsilon$ does not precisely correspond to the WBLA. In contrast, for the standard GKBA approach, at different levels of approximation, the corresponding result changes with the number of poles in the expansion~\eqref{eq:pole:exp}. The main panels in Fig.~\ref{fig:weak} include $N_p=30$ poles, and the pole count is varied at the inset of panel Fig.~\ref{fig:weak}(c), showing a slow divergence (logarithmic) in the case of WBLA. Interestingly, the GKBA calculations with finite-width Lorentzians do converge and the convergence is relatively fast, around $N_p \gtrsim 10$ poles, but they converge to an unphysical value ($\neq 0$). The overall mismatch in the energy currents is still fairly minimal at weak coupling.

\begin{figure}[t]
\center
\includegraphics[width=0.95\columnwidth]{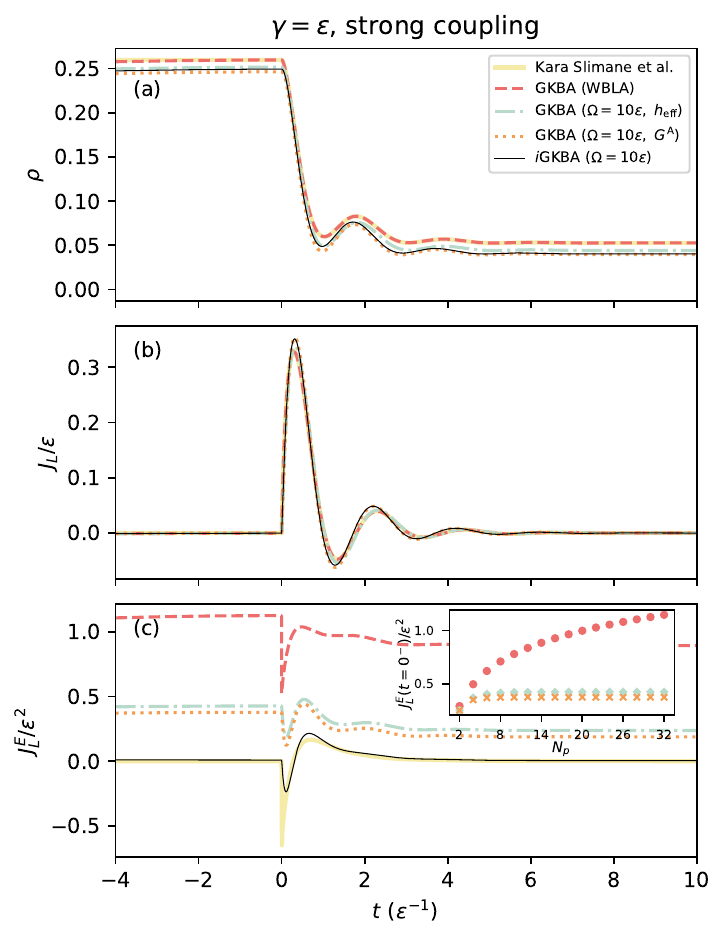}
\begin{caption}{
Same as Fig.~\ref{fig:weak} but the quantum dot being strongly coupled to the leads, $\gamma=\varepsilon$.
\label{fig:strong}}
\end{caption}
\end{figure}

The $i$GKBA scheme accurately resolves the energy currents in the strong coupling regime as well; see Fig.~\ref{fig:strong}. Here, the parameters are otherwise the same as in the weak-coupling calculation (Fig.~\ref{fig:weak}) but the coupling strength is varied so that $\gamma=\varepsilon$. Interestingly, the charge density and current [Fig.~\ref{fig:strong}(a) and Fig.~\ref{fig:strong}(b)] are accurately described by the standard GKBA, but the energy current again diverges for the WBLA case with increasing pole count. We clearly see that when the coupling strength is increased, the standard GKBA at the level of WBLA becomes entirely insufficient. The GKBA calculations with finite-band Lorentzian converge again around $N_p\gtrsim 10$ poles, but there is still a noticeable shift in the equilibrium energy current. The $i$GKBA corrects this unphysical behavior, but it is worth noting that the finite-bandwidth Lorentzian with $\Omega=10\varepsilon$ description deviates more from the result of Ref.~\cite{kara-slimane_simulating_2020}, obtained at the WBLA, when the coupling is stronger. Still, it is required to go beyond GKBA to fix the problem of unphysical energy currents in equilibrium.

\begin{figure}[t]
\center
\includegraphics[width=0.95\columnwidth]{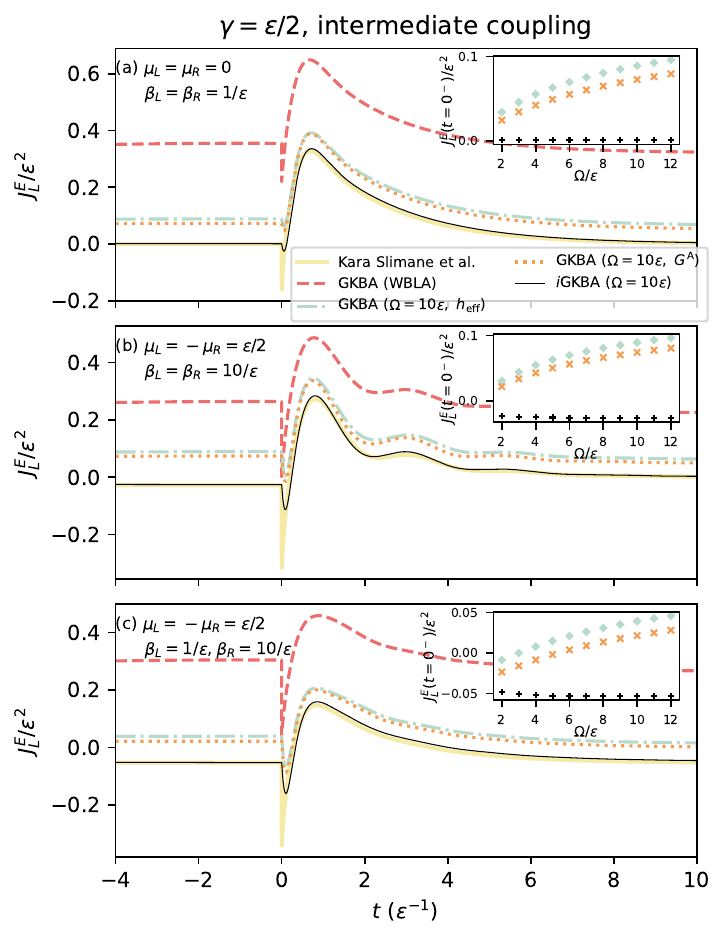}
\begin{caption}{
Time-dependent energy current $J_L^E$ at the left-lead interface with the quantum dot being coupled to the leads with $\gamma=\varepsilon/2$, and driven by a sudden gate voltage $u(t)=(5\varepsilon/2)\theta(t)$. (a) High temperature $\mu_L=\mu_R=0$, $\beta_L=\beta_R=1/\varepsilon$; (b) chemical potential drop $\mu_L=-\mu_R=\varepsilon/2, \beta_L=\beta_R=10/\varepsilon$; and (c) temperature gradient $\mu_L=-\mu_R=\varepsilon/2, \beta_L=1/\varepsilon, \beta_R=10/\varepsilon$. Insets display the equilibrium energy current, before the sudden gate voltage is switched on, in terms of the Lorentzian bandwidth $\Omega$.
\label{fig:inter}}
\end{caption}
\end{figure}

With charge densities and currents showing excellent agreement across different levels of approximation and with the earlier literature, we turn to the energy-current response in the intermediate coupling regime $\gamma=\varepsilon/2$, modifying the transport setup accordingly. Fig.~\ref{fig:inter} displays three different cases (a) aligned chemical potentials $\mu_L=\mu_R=0$ at high temperature $\beta_L=\beta_R=1/\varepsilon$; (b) chemical potential drop $\mu_L=-\mu_R=\varepsilon/2$ at lower temperature $\beta_L=\beta_R=10/\varepsilon$; and (c) chemical potential drop $\mu_L=-\mu_R=\varepsilon/2$ with temperature gradient $\beta_L=1/\varepsilon, \beta_R=10/\varepsilon$. In all cases, the pole count is set to $N_p=40$. The GKBA (WBLA) result is again shifted the most from the equilibrium value. For GKBA, the finite-width Lorentzians capture the correct trend well in all cases, but there is always a shift in the equilibrium value. It is worth pointing out that now the cases in Fig.~\ref{fig:inter}(b)~and~\ref{fig:inter}(c) display chemical potential drops and thermal gradients, i.e., the equilibrium energy current being non-zero is physically expected. In all cases, the $i$GKBA approach corrects for the unphysical behavior observed in GKBA. In the insets of Fig.~\ref{fig:inter} we also show the dependence on the Lorentzian bandwidth $\Omega$. For the equilibrium energy current, the GKBA results are not only unphysical but the convergence toward the WBLA at $\Omega\to\infty$ is fairly slow. In contrast, the $i$GKBA approach accurately resolves the equilibrium energy current, which does not vary significantly with the bandwidth $\Omega$.

\begin{figure}[t]
\center
\includegraphics[width=0.95\columnwidth]{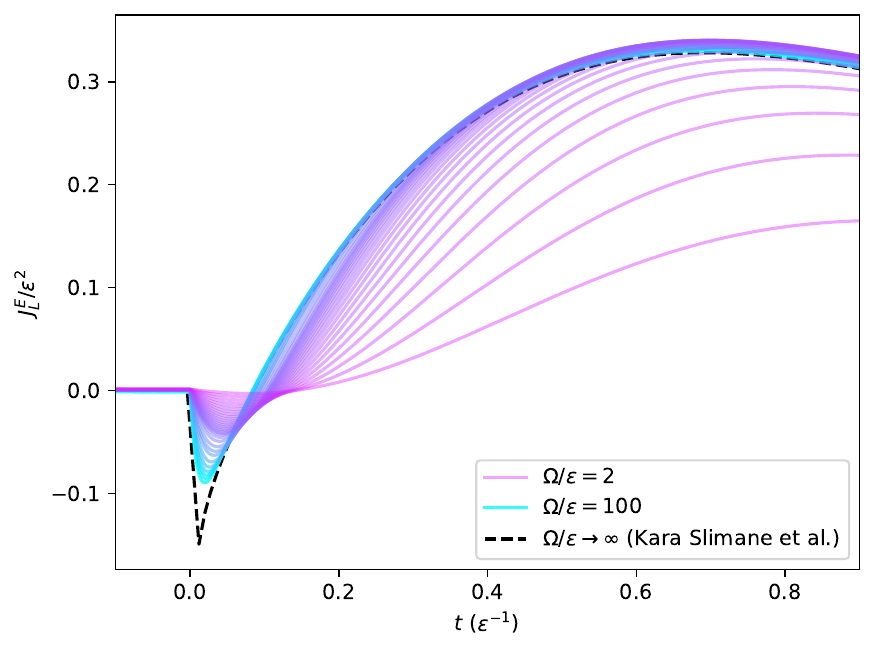}
\begin{caption}{
Short-time transient behavior of the energy current. The WBLA result displays an abrupt jump at the voltage switch-on time, which can be continuously modeled by applying wider bandwidth Lorentzians. The broadening parameter $\Omega/\varepsilon\in[2,100]$ is shown in varying color, from magenta to cyan.
\label{fig:initial}}
\end{caption}
\end{figure}

The broadening parameter $\Omega$ in the finite-width Lorentzian has a pronounced impact on the initial transient dynamics, as already evidenced in Figs.~\ref{fig:weak}-\ref{fig:inter}. To further illustrate this behavior, Fig.~\ref{fig:initial} zooms into the early-time regime of the energy current shown in Fig.~\ref{fig:inter}(a), immediately following the gate-voltage quench at $t=0$. With the analytic results at the WBLA~\cite{kara-slimane_simulating_2020}, we see that the energy current exhibits an abrupt jump at the time of the voltage switch-on. The jump can be shown to build up continuously by applying a larger broadening $\Omega$ in the $i$GKBA scheme. Similar findings have also been reported in the case of thermomechanical potentials across the junction~\cite{covito_transient_2018}. At larger times $t\gtrsim \varepsilon^{-1}$, the energy currents calculated via large-bandwidth Lorentzians again resolve the WBLA result accurately.

\section{Comparison of ($i$)GKBA with the full Kadanoff-Baym equations}
We now consider an interacting molecular junction modeled at the Pariser-Parr-Pople~\cite{pariser_semi-empirical_1953, pople_electron_1953} (PPP) level, where the kinetic and interaction matrix elements are semi-empirically obtained by fitting to more sophisticated calculations. Specifically, we study a cyclobutadiene molecule attached to donor-acceptor-like leads~\cite{tuovinen_electronic_2021}. This molecule consists of four atomic sites arranged in a ring (Fig.~\ref{fig:system}) and is modeled using PPP parameters derived from an effective valence shell Hamiltonian~\cite{martin_ab-initio_1996}. Referring to Eq.~\eqref{eq:ham}, the single-particle matrix (in atomic units) is given by
\begin{equation}
h=-\begin{pmatrix} 0.90286 & 0.11908 & 0 & 0.09772 \\
  	  	         0.11908 & 0.90286 & 0.09772 & 0 \\
		           0  & 0.09772 & 0.90286 & 0.11908 \\
		           0.09772 & 0 & 0.11908 & 0.90286 \end{pmatrix},
\end{equation}
and the two-body interaction is of the form $v_{mnpq}=v_{mn}\delta_{mq}\delta_{np}$, with (in atomic units)
\begin{equation}
v = \begin{pmatrix} 0.43255 & 0.20143 & 0.16515 & 0.20181 \\
            		0.20143 & 0.43048 & 0.20181 & 0.16515 \\
            		0.16515 & 0.20181 & 0.43255 & 0.20143 \\
            		0.20181 & 0.16515 & 0.20143 & 0.43048 \end{pmatrix} .
\end{equation}
The slight asymmetry in the hopping and interaction matrix elements arises from the fact that cyclobutadiene forms a rectangular rather than a square structure~\cite{breslow_charge_2008}. We note in passing that the physico-chemical properties of the cyclobutadiene molecule are fairly multifaceted~\cite{rosenberg_excited_2014}. Our aim here is to assess the validity of the $i$GKBA scheme when compared against the Kadanoff-Baym equations, independent of how accurate the PPP model description is, and we turn aside further discussions on, e.g., the antiaromatic nature and the resulting instability of the molecule~\cite{hong_manifestations_2022}.

The coupling between the molecule and the leads is symmetric: the coupling strength between the molecular sites $1$ and $4$ and the left lead is equal to that between sites $2$ and $3$ and the right lead: 
\begin{equation}
\gamma_{\a,ij}=\delta_{ij}\left[\gamma_L^0\delta_{\a L}(\delta_{i1}+\delta_{i4}) + \gamma_R^0\delta_{\a R}(\delta_{i2}+\delta_{i3})\right]
\end{equation}
with $\gamma_L^0=\gamma_R^0\equiv\gamma^0/4$. We note that $\gamma_\a$ is a matrix while $\gamma_{(\alpha)}^0$ are scalars indicating the coupling strength.
To define energy scales, we use the difference between the highest occupied molecular orbital (HOMO) and the lowest unoccupied molecular orbital (LUMO): $\Delta \equiv E_{\mathrm{LUMO}}-E_{\mathrm{HOMO}} = 0.244151$~a.u., calculated at the Hartree-Fock level. For charge neutrality in equilibrium, the chemical potential of the lead-molecule system is set in the middle of the HOMO-LUMO gap: $\mu=-0.118715$~a.u., corresponding to the isolated molecule with two electrons. We consider weak ($\gamma^0=\Delta/10$) and strong coupling ($\gamma^0=\Delta$), as well as low and high temperatures, defined by $\beta=40/\Delta$ and $\beta=8/\Delta$, respectively. Note that the `low' temperature is still relatively high when converted to physical units, but here, thermal broadening begins to affect the system only when $k_{\mathrm{B}}T=1/\beta$ becomes comparable to the molecular energy levels.

The molecular junction is driven out of equilibrium by a sudden bias voltage on the leads, $V_\a(t)=V_\a^0\theta(t)$, and we consider weak ($V_L^0=-V_R^0=\Delta/8$) and strong ($V_L^0=-V_R^0=\Delta/2$) bias voltages. In these situations, the energy centroids of the leads are set to $\epsilon_{L/R}=\mu\pm\Delta/8$ for the weak driving and $\epsilon_{L/R}=\mu\pm\Delta/2$ for the strong driving. The driving parameters are chosen to highlight the capability of the $i$GKBA scheme~\cite{pavlyukh_open_2025}, which does not rely on the uniformity of the lead density of states. The strong driving case is enough to excite a sizable electric current across the HOMO-LUMO gap.

We include the effect of two-body interaction at the Hartree-Fock level, i.e., we neglect the correlation self-energy in Eq.~\eqref{eq:gf-eom}. This simplified modeling of the molecular junction allows us to explore the new approach with mathematical transparency, facilitating the comparison of the $i$GKBA approach with the full Kadanoff-Baym equations.

\begin{figure}[t]
\center
\includegraphics[width=0.95\columnwidth]{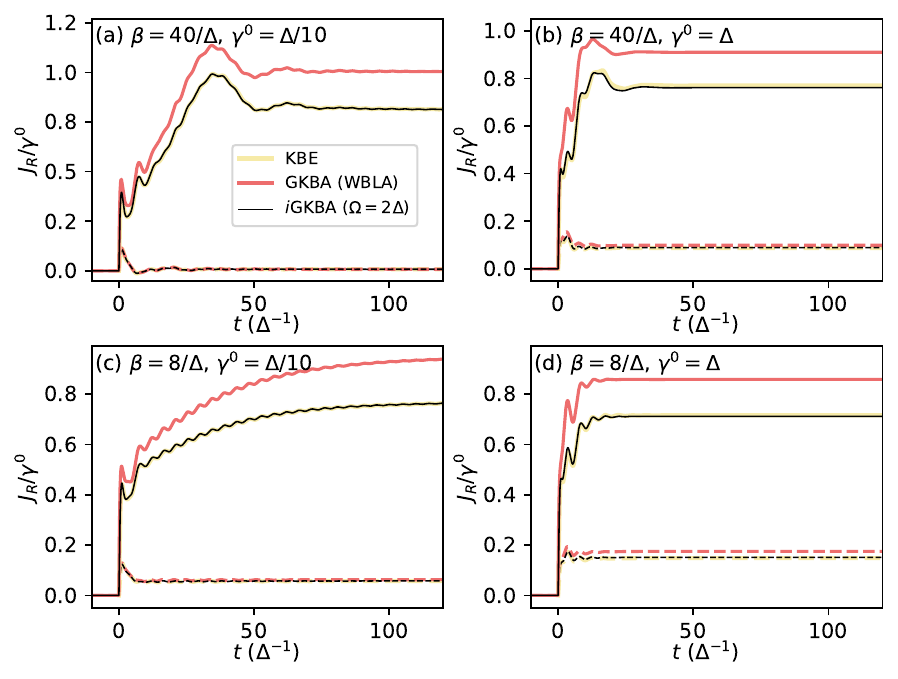}
\begin{caption}{
Time-dependent charge current at the right-lead interface in the cyclobutadiene molecular junction: (a) weak coupling, low temperature; (b) strong coupling, low temperature; (c) weak coupling, high temperature; (d) strong coupling, high temperature. Solid and dashed lines correspond to the strong and weak driving cases, respectively.
\label{fig:kb-comparison}}
\end{caption}
\end{figure}

In Fig.~\ref{fig:kb-comparison}, we present comparative simulations of the various methods for the time-dependent charge current at the right-lead interface. The benchmark KBE results are based on the method of Ref.~\cite{myohanen_kadanoff-baym_2009}, which we supplement with the Lorentzian line-width function for the embedding self-energies, in order to consistently compare it with the present development. We take the finite-bandwidth Lorentzians as $\Omega=2\Delta$. In contrast to the adiabatic switching procedure in the ($i$)GKBA methods, the KBE approach determines the coupled equilibrium state by solving the Dyson equation on the imaginary time axis~\cite{dahlen_self-consistent_2005, balzer_nonequilibrium_2009-1, tuovinen_adiabatic_2019}. In practice, the imaginary time grid, $\tau\in[0,-\beta]$, is discretized using a uniform power mesh~\cite{ku_band-gap_2002, schuler_spectral_2018} with $p=6$, $u=8$ (total number of $\tau$-points being $2pu+1$) to achieve convergence in the total energy up to the relative error $10^{-6}$ for all the cases considered.

In all the cases reported in Fig.~\ref{fig:kb-comparison}, we see that $i$GKBA systematically improves upon the GKBA result, and the agreement with KBE is mostly excellent. While the initial transient response to the sudden bias voltage and the following oscillations are qualitatively similar in all methods, quantitative differences appear. The most prominent improvements of the $i$GKBA are the slow decay toward the stationary state and the value of the stationary current. These are both incorrectly described by the GKBA, especially in the strong-bias cases.

\section{Time-resolved thermoelectric energy conversion}

Now that the validity range of the $i$GKBA approach has been established, we continue with thermoelectric simulations for the molecular junction of the previous section. In addition to a bias voltage, we apply a temperature gradient with $\beta_L\neq\beta_R$ over the junction. To address thermoelectric energy conversion, we look at the total charge and heat currents $J^{(H)}_{\mathrm{tot}}=(J^{(H)}_R-J^{(H)}_L)/2$ with $J^H_\alpha = J^E_\alpha - \mu_\alpha J_\a$. In the linear-response regime, the relation between currents and driving fields can be expressed in matrix form as~\cite{galperin_inelastic_2008a, heikkila_thermal_2019}
\begin{equation}\label{eq:coefficients}
\begin{pmatrix} J_{\mathrm{tot}} \\ J^H_{\mathrm{tot}} \end{pmatrix} = \begin{pmatrix} \mathcal{G} & \mathcal{L} \\ \mathcal{R} & \mathcal{K} \end{pmatrix} \begin{pmatrix} \delta V \\ \delta T \end{pmatrix}, 
\end{equation}
where $\mathcal{G},\mathcal{L},\mathcal{R},\mathcal{K}$ denote Onsager's kinetic coefficients. The physical transport coefficients, such as electrical and thermal conductivities, and the Seebeck and Peltier coefficients, are defined in terms of these~\cite{callen_application_1948}. From Eq.~\eqref{eq:coefficients} we obtain, in the absence of charge current~\cite{crepieux_enhanced_2011},
\begin{equation}\label{eq:seebeck}
S\equiv -\left.\frac{\delta V}{\delta T}\right|_{J_{\mathrm{tot}}=0} = \frac{\mathcal{L}}{\mathcal{G}},
\end{equation}
which defines the Seebeck coefficient, or thermopower. This relation then enables the calculation of time-resolved thermopower by adjusting the bias voltage so that the charge current vanishes, thereby providing a basis for characterizing thermoelectric energy conversion in the molecular junction under non-stationary conditions.

For a more realistic driving protocol, we apply a time-dependent voltage profile, $V_\alpha(t) = V_\alpha^0 / [1+\exp(-\kappa t)]$ with $\kappa=5\Delta$, which means that around $t=0$ the bias rapidly yet smoothly increases from $0$ to $V_\alpha^0$. Applying a thermal gradient over the junction, $\delta T\equiv 1/\beta_L - 1/\beta_R>0$, generates a non-zero charge current, which can be `countered' by applying a voltage $\delta V\equiv V_L^0 - V_R^0<0$. In practice, this type of thermoelectric energy harvesting involves driving an electric current against a load using a thermal gradient as a driving force~\cite{sothmann_thermoelectric_2015}. Our motivation here is for dynamically enhanced thermoelectric efficiency in a far-from-equilibrium setting, possibly surpassing steady-state limitations set by static material properties and compatibility conditions~\cite{snyder_thermoelectric_2003, benenti_fundamental_2017, zhou_boosting_2015}.

\begin{figure}[t]
\center
\includegraphics[width=0.95\columnwidth]{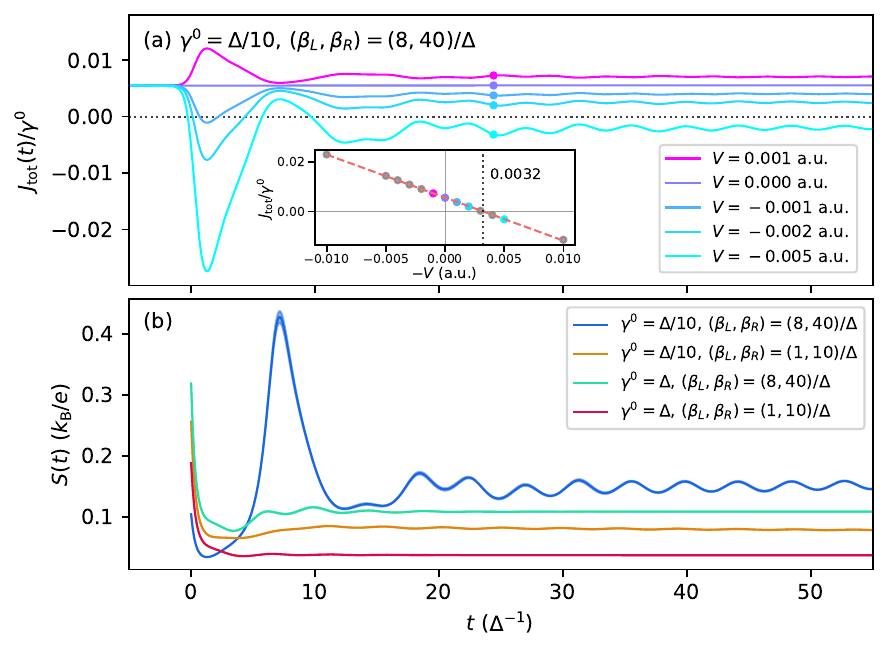}
\begin{caption}{
Time-resolved thermoelectricity in the cyclobutadiene molecular junction. (a) Total charge currents through the molecular junction with $\gamma^0=\Delta/10$ and $(\beta_L,\beta_R)=(8,40)/\Delta$, for varying applied voltages $V\equiv V_L^0=-V_R^0$. The inset shows instantaneous current-voltage characteristics indicated by the markers in the main panel; gray markers correspond to calculations not shown in the main panel. A linear fit is employed for obtaining the intercept voltage that suppresses the charge current, indicated by the vertical dotted line. (b) Extracted thermopower for different values of the coupling strength and lead temperatures.
\label{fig:thermopower}}
\end{caption}
\end{figure}

The procedure for finding the time-resolved thermopower is visualized in Fig.~\ref{fig:thermopower}(a), where the total charge current through the molecular junction is shown with varying voltages. Notably, at the stationary state, there is a non-zero charge current because of the thermal gradient only, without applied voltage. At each instant of time, we collect the charge current values and plot them against the applied voltage, and we observe that these values systematically follow a linear relationship [inset in Fig.~\ref{fig:thermopower}(a)]. Performing a linear fit for all steps during the time evolution, we can extract the voltage value at which the charge current vanishes. Although the applied voltage must `counter' ($\delta V<0$) the thermal gradient in order to suppress the charge current, we also include positive voltages for a more accurate fit. According to Eq.~\eqref{eq:seebeck}, this procedure then gives us the time-resolved thermopower.

The time-resolved thermopower is shown in Fig.~\ref{fig:thermopower}(b) for different values of the coupling strength $\gamma^0$ and the lead temperatures $(\beta_L,\beta_R)$. During the adiabatic switching period ($t<0$), in the absence of applied voltages, the charge currents have saturated to certain non-zero values. The thermopower is then evaluated for times $t>0$ only. It exhibits a rapid drop within a few $\Delta^{-1}$ due to fast charge fluctuations in the molecule. For $\gamma^0=\Delta/10$ and $(\beta_L,\beta_R)=(8,40)/\Delta$, we see a clear transient enhancement of the thermopower, similar to the one reported in Ref.~\cite{crepieux_enhanced_2011}. The transient behavior thus indicates an increased efficiency of the conversion of thermoelectric energy. Since the parameter space of our thermoelectric transport setup is very large, we do not aim to study exhaustively the optimal conditions for this effect but defer this to future work. In addition, the determined thermopower curves include standard deviations originating from the linear-fit procedure explained above, and we observe that the linear relationship holds very accurately throughout the transient dynamics. Similar to the extraction of the time-resolved thermopower, also the thermoelectric figure of merit, $ZT\equiv\mathcal{G}S^2T/\mathcal{K}$, may be obtained from a similar analysis based on the charge and heat current data~\cite{suppmat}.

\begin{figure}[t]
\center
\includegraphics[width=0.95\columnwidth]{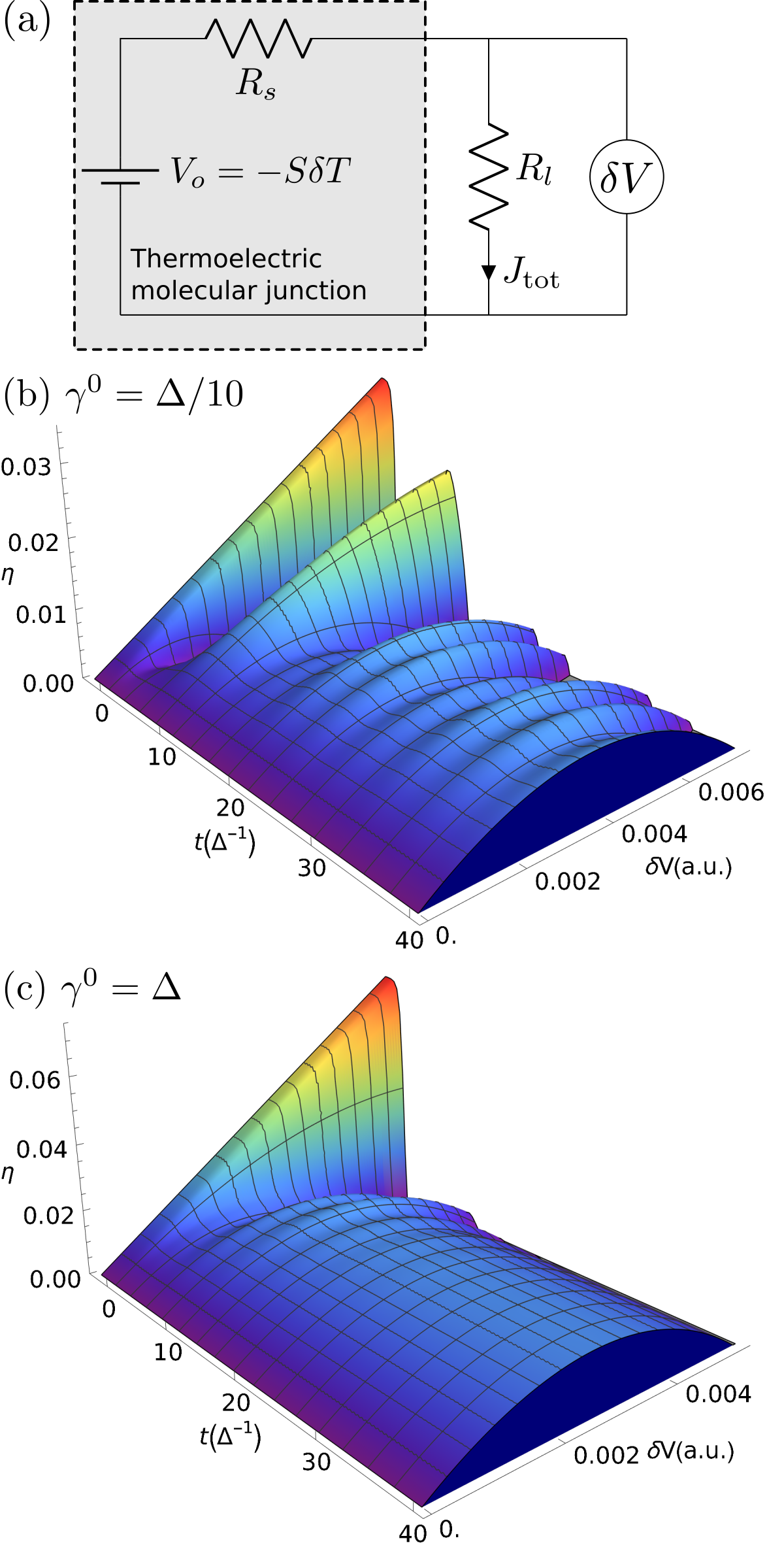}
\begin{caption}{
Thermoelectric energy conversion efficiency in the cyclobutadiene molecular junction. (a) Circuit schematic for the thermoelectric molecular junction connected to an external load; (b,c) time- and voltage-resolved efficiency for coupling strengths $\gamma^0 = \{\Delta / 10, \Delta\}$, respectively.   
\label{fig:efficiency}}
\end{caption}
\end{figure}

Finally, we study the efficiency of the thermoelectric molecular junction functioning as a heat engine. This can be achieved using the following construction in the linear-response and stationary regimes [see Eq.~\eqref{eq:coefficients} and Fig.~\ref{fig:efficiency}(a)]: The thermoelectric molecular junction (hereafter referred to as \emph{the system}) is connected to an external load and thermalized in the presence of a temperature gradient. Since the system operates under a temperature gradient $\delta T$, it develops an open-circuit Seebeck voltage $V_o = -S\delta T$, as extracted in Fig.~\ref{fig:thermopower}. The system has an internal resistance $R_s$ and is connected in series with an external load of resistance $R_l$. This configuration forms a closed circuit supporting a total current $J_{\mathrm{tot}} = V_o / (R_s + R_l)$. 

There are two limiting cases:  
(1) A short-circuit situation arises during the thermalization when $R_l = 0$, in which case the voltage at the load is $\delta V = J_{\mathrm{tot}} R_l = 0$. The corresponding short-circuit current is $J_s = V_o / R_s$.  
(2) An open-circuit situation occurs when $R_l \to \infty$, in which case the voltage at the load equals the Seebeck voltage, $\delta V = V_o$, and the current vanishes, as already seen in Fig.~\ref{fig:thermopower}. 

In intermediate cases, the total current takes the form $J_{\mathrm{tot}} = J_s - \delta V / R_s$ (by Kirchhoff's junction rule), from which we infer the power output at the load: $J_{\mathrm{tot}}\delta V = (J_s - \delta V / R_s)\delta V$. In the stationary state, the power thus follows a downward-opening parabola in $\delta V$, with a maximum at $\delta V = V_o / 2$, corresponding to the impedance-matching condition $R_l = R_s$.

The energy conversion efficiency of the thermoelectric junction is then given by
\begin{equation}\label{eq:efficiency}
\eta(t) = \frac{J_{\mathrm{tot}}(t)\delta V}{J_L^H(t)},
\end{equation}
where $J_L^H$ is the heat current from the left (hot) lead to the molecule. This formulation allows us to evaluate $\eta$ without requiring explicit knowledge of either $R_s$ or $R_l$. Instead, we parametrize the system in terms of the observable load voltage $\delta V$, which can be held approximately constant even in the transient regime by connecting a capacitor in parallel to the load, and which we scan over the interval $[0, V_o]$. Due to the time-dependent nature of the setup, we set $V_o$ equal to the mean value of the extracted Seebeck voltages shown in Fig.~\ref{fig:thermopower}~\cite{suppmat}. This captures both the short- and open-circuit limits, where the efficiency necessarily vanishes due to zero extracted power. The efficiency reaches a maximum at an intermediate value of $\delta V$, allowing a full characterization of the junction performance based solely on measurable quantities.

We focus on two representative cases by fixing the temperature difference as $(\beta_L, \beta_R) = (8, 40)/\Delta$ and varying the coupling strength $\gamma^0 = \{\Delta / 10, \Delta\}$, for which the extracted mean values of the Seebeck voltages are $V_o = \{0.0075, 0.0053\}$~a.u., respectively~\cite{suppmat}. Using Eq.~\eqref{eq:efficiency}, the time- and voltage-resolved efficiency is shown in Fig.~\ref{fig:efficiency}(b,c). Toward the stationary state, the efficiency takes a parabolic dependence on the load voltage $\delta V$ as expected. In both coupling cases, the efficiency initially grows linearly with $\delta V$ during the transient but eventually saturates to the parabolic shape. The saturation time is predictably longer in the weak-coupling case, where the thermoelectric efficiency exhibits periodic enhancements over tens of $\Delta^{-1}$. With the same parameters, a similarly enhanced thermopower was observed in Fig.~\ref{fig:thermopower}(b). For stronger coupling [Fig.~\ref{fig:efficiency}(c)], the transient oscillations are suppressed, and within $t \sim 10 \Delta^{-1}$ the expected parabolic behavior sets in.

We observe that the transiently enhanced efficiency can clearly surpass the stationary value. It should be noted, however, that the calculated efficiencies are relatively low ($\lesssim 10$\%), which is attributed to the specific molecular structure, the coupling configuration, and the applied temperature gradient. For voltages outside the range $[0, V_o]$, energy is effectively pumped into the system, and the efficiency drops significantly or becomes negative, as dictated by the direction of currents and the form of Eq.~\eqref{eq:efficiency}. More optimized conditions for efficient thermoelectric energy conversion could be explored within the parameter space of this model, or by considering alternative molecular junction setups, and we defer more thorough investigations of quantum thermodynamics phenomena~\cite{esposito_quantum_2015, michelini_entropy_2017} to future work. Ultimately, the Carnot efficiency $\eta_C=1 - T_R / T_L$ sets the fundamental upper limit. This bound is only reached in the idealized limit where the thermoelectric figure of merit $ZT\to\infty$; for any finite $ZT$, the efficiency remains strictly below $\eta_C$~\cite{benenti_thermodynamic_2011,muralidharan_performance_2012,suppmat}.

\section{Conclusion}

We investigated thermoelectric dynamics in open quantum systems beyond the wide-band limit, specifically focusing on our recently-developed iterated generalized Kadanoff-Baym ansatz~\cite{pavlyukh_open_2025}. This entailed a time-linear scaling, non-equilibrium Green's function theory for fast and accurate simulation of open system dynamics, where the environment has a non-flat spectral density, in contrast with the standard wide-band limit approximation. Our approach of iterating the reconstruction equation for the lesser and greater Green's function includes non-Markovian effects, where the interaction with the environment exhibits memory-dependent behavior.

Applying the GKBA iteratively in the reconstruction equation resulted in a set of 14 coupled equations of motion, referred to as the $i$GKBA scheme. Starting from the full $i$GKBA scheme, different levels of approximation were explored. Neglecting correction terms in the collision integral reduced to the GKBA framework, still incorporating retarded/advanced propagators beyond the wide-band limit. We exemplified the different levels of approximation by focusing on thermoelectric transport in a two-terminal quantum dot device, for which analytical benchmark results were available. The conventional GKBA proved insufficient for energy current calculations due to an unphysical divergence in the WBLA, which was resolved by the finite-bandwidth $i$GKBA scheme. We also found that the short-time transient behavior of the energy current exhibited an abrupt jump, which was correctly resolved by wider-bandwidth Lorentzians in the $i$GKBA scheme.

Finally, we modeled thermoelectric transport in a cyclobutadiene molecular junction --- an interacting system --- analyzing the non-equilibrium dynamics under both weak and strong coupling, as well as varying temperatures. By comparing with the numerical solution of the full Kadanoff-Baym equations, the $i$GKBA scheme was found to systematically improve upon GKBA by more accurately capturing relaxation dynamics and steady-state currents. Using the transient charge and heat currents, we investigated thermoelectric energy conversion by extracting the time-resolved Seebeck coefficient (thermopower) and the device efficiency. The molecular junction, when operating in the transient regime, exhibited a clear enhancement in thermoelectric conversion efficiency compared to its stationary-state performance.

In summary, our approach advances quantum simulations of time-resolved thermoelectric phenomena, offering accurate and efficient access to the ultrafast dynamics of charge and heat flow at the nanoscale. By overcoming limitations of traditional methods, this bottom-up approach opens the door to predictive modeling of, e.g., transistor technology using two dimensional materials~\cite{yoon_how_2011, pizzi_performance_2016} and ultrafast bolometers for qubit readout in quantum computing~\cite{gunyho_single-shot_2024}. Our framework is well-suited to address the complex, time-dependent transport phenomena and energy-efficient device design.

\acknowledgments
R.T. acknowledges the financial support of the Jane and Aatos Erkko Foundation (Project EffQSim) and the Research Council of Finland through the Finnish Quantum Flagship (Project No. 359240).
We also acknowledge CSC--IT Center for Science, Finland, for computational resources.


%

\widetext
\newpage

\begin{center}
\textbf{\large Supplemental Material: Thermoelectric energy conversion in molecular junctions out of equilibrium}
\end{center}
\setcounter{equation}{0}
\setcounter{figure}{0}
\setcounter{table}{0}
\setcounter{page}{1}
\setcounter{section}{0}
\makeatletter
\renewcommand{\theequation}{S\arabic{equation}}
\renewcommand{\thefigure}{S\arabic{figure}}
\renewcommand{\thepage}{S\arabic{page}}
\renewcommand{\thesection}{S\arabic{section}}
\renewcommand{\bibnumfmt}[1]{[S#1]}
\renewcommand{\citenumfont}[1]{S#1}
\def\tb{\bar{t}}
\def\im{{i}}
\def\ex{{{e}}}
\def\ud{{{d}}}
\def\heff{h_{\text{eff}}}
\def\unity{\mathds{1}}

\bigskip\bigskip

\section{Energy-current formula in the wide-band limit}

The embedding self-energies related to charge ($\nu=0$) and energy ($\nu=1$) currents are
\begin{equation}
{\Sigma}_{\a,mn}^{\nu,<}(t,t') = \sum_k E_{k\a}^\nu T_{mk\a}s_\a(t)g_{k\a}^<(t,t')T_{k\a n}s_\a (t'),
\end{equation}
where $s_\a$ is the ramp function for the contacts, and the free Green's function of the leads' is
\begin{equation}
g_{k\a}^<(t,t') = \im f_\a(E_{k\a}-\mu_\a)\ex^{-\im\int_{t'}^t \ud \tb [E_{k\a}+V_\a(\tb)]}.
\end{equation}
We focus on the energy current case with $\nu=1$. By using the short-hand notation for the bias-voltage phase factor $\phi_\a(t,t') \equiv \int_{t'}^t \ud \tb V_\a(\tb)$, and by converting the $k$-summation into an integral we obtain
\begin{align}\label{eq:convert}
{\Sigma}_{\a,mn}^{1,<}(t,t') & = \im \ex^{-\im\phi_\a(t,t')}\sum_k \int_{-\infty}^\infty\frac{\ud\w}{2\pi} 2\pi\delta(\w-E_{k\a})\w T_{mk\a}s_\a(t)f_\a(\w-\mu_\a)\ex^{-\im\w(t-t')}T_{k\a n}s_\a(t')\nonumber \\
& = \im s_\a(t) s_\a(t') \ex^{-\im\phi_\a(t,t')} \int_{-\infty}^\infty \frac{\ud\w}{2\pi} \w f_\a(\w-\mu)\ex^{-\im \w(t-t')} \underbrace{2\pi \sum_k T_{mk\a}\delta(\w-E_{k\a})T_{k\a n}}_{\equiv \Gamma_{\a,mn}(\w)}.
\end{align}
The line-width function becomes independent of frequency, $\Gamma_\a(\w) \approx \gamma_\a$, in the wide-band limit approximation (WBLA). Using the pole expansion of the Fermi function from the main text Eq.~(21) we find
\begin{align}
\Sigma_{\a}^{1,<}(t,t') & = \im\gamma_\a s_\a(t)s_\a(t')\ex^{-\im\phi_\a(t,t')}\frac{1}{2}\int_{-\infty}^\infty\frac{\ud\w}{2\pi}\w \ex^{-\im\w(t-t')} \nonumber \\
& - \im\gamma_\a s_\a(t)s_\a(t')\ex^{-\im\phi_\a(t,t')}\sum_\ell \frac{\eta_\ell}{\beta_\a}\int_{-\infty}^\infty\frac{\ud\w}{2\pi}\w\left(\frac{1}{\w-\mu_\a+\im\zeta_\ell/\beta_\a}+\frac{1}{\w-\mu_\a-\im\zeta_\ell/\beta_\a}\right)\ex^{-\im\w(t-t')}.
\end{align}
For the first line, a representation of the Dirac delta function can be utilized:
\begin{equation}
\delta(t-t') = \int_{-\infty}^\infty \frac{\ud\w}{2\pi}\ex^{-\im\w(t-t')} \quad \Rightarrow \quad -\im\frac{\partial}{\partial t'} \delta(t-t') = \int_{-\infty}^\infty \frac{\ud\w}{2\pi} \w\ex^{-\im\w(t-t')}.
\end{equation}
For the second line, a manipulation is performed:
\begin{align}
& \int_{-\infty}^\infty\frac{\ud \w}{2\pi}\w\left(\frac{1}{\w-\mu_\a+\im\zeta_\ell/\beta_\a}+\frac{1}{\w-\mu_\a-\im\zeta_\ell/\beta_\a}\right)\ex^{-\im\w(t-t')} \nonumber \\
& = \int_{-\infty}^\infty\frac{\ud \w}{2\pi}\frac{\w}{\w-\mu_\a+\im\zeta_\ell/\beta_\a}\ex^{-\im\w(t-t')} + \int_{-\infty}^\infty\frac{\ud \w}{2\pi}\frac{\w}{\w-\mu_\a-\im\zeta_\ell/\beta_\a} \ex^{-\im\w(t-t')} \nonumber \\
& = \int_{-\infty}^\infty\frac{\ud \w}{2\pi}\frac{\w-\mu_\a+\im\zeta_\ell/\beta_\a+\mu_\a-\im\zeta_\ell/\beta_\a}{\w-\mu_\a+\im\zeta_\ell/\beta_\a}\ex^{-\im\w(t-t')} + \int_{-\infty}^\infty\frac{\ud \w}{2\pi}\frac{\w-\mu_\a-\im\zeta_\ell/\beta_\a+\mu_\a+\im\zeta_\ell/\beta_\a}{\w-\mu_\a-\im\zeta_\ell/\beta_\a} \ex^{-\im\w(t-t')} \nonumber \\
& = \int_{-\infty}^\infty\frac{\ud \w}{2\pi}\left(1+\frac{\mu_\a-\im\zeta_\ell/\beta_\a}{\w-\mu_\a+\im\zeta_\ell/\beta_\a}\right)\ex^{-\im\w(t-t')} + \int_{-\infty}^\infty\frac{\ud \w}{2\pi}\left(1+\frac{\mu_\a+\im\zeta_\ell/\beta_\a}{\w-\mu_\a-\im\zeta_\ell/\beta_\a}\right) \ex^{-\im\w(t-t')}. \nonumber
\end{align}
This expression can then be separated and evaluated using the Cauchy integral formula:
\begin{align}
& \int_{-\infty}^\infty\frac{\ud \w}{2\pi}\left(1+\frac{\mu_\a-\im\zeta_\ell/\beta_\a}{\w-\mu_\a+\im\zeta_\ell/\beta_\a}\right)\ex^{-\im\w(t-t')} + \int_{-\infty}^\infty\frac{\ud \w}{2\pi}\left(1+\frac{\mu_\a+\im\zeta_\ell/\beta_\a}{\w-\mu_\a-\im\zeta_\ell/\beta_\a}\right) \ex^{-\im\w(t-t')} \nonumber \\
& = \int_{-\infty}^\infty\frac{\ud \w}{2\pi}\ex^{-\im\w(t-t')} + (\mu_\a-\im\zeta_\ell/\beta_\a)\int_{-\infty}^\infty\frac{\ud \w}{2\pi}\frac{1}{\w-\mu_\a+\im\zeta_\ell/\beta_\a}\ex^{-\im\w(t-t')} \nonumber \\
& + \int_{-\infty}^\infty\frac{\ud \w}{2\pi}\ex^{-\im\w(t-t')} + (\mu_\a+\im\zeta_\ell/\beta_\a)\int_{-\infty}^\infty\frac{\ud \w}{2\pi}\frac{1}{\w-\mu_\a-\im\zeta_\ell/\beta_\a} \ex^{-\im\w(t-t')} \nonumber \\
& = \left[\delta(t-t')-\im(\mu_\a-\im\zeta_\ell/\beta_\a)\ex^{-\im(\mu_\a-\im\zeta_\ell/\beta_\a)(t-t')}\theta(t-t') +\delta(t-t')+\im(\mu_\a+\im\zeta_\ell/\beta_\a)\ex^{-\im(\mu_\a+\im\zeta_\ell/\beta_\a)(t-t')}\theta(t'-t)\right].
\end{align}
For the time-convolutions of the embedding self-energy [main text Eq.~(9)], only the $t>t'$ contribution is required:
\begin{align}
{\Sigma}_{\a}^{1,<}(t,t') & = \frac{1}{2}s_\a(t)s_\a(t')\ex^{-\im\phi_\a(t,t')}\frac{\partial}{\partial t'} \delta(t-t')\gamma_a - 2\im s_\a(t)s_\a(t')\ex^{-\im\phi_\a(t,t')}\sum_\ell \frac{\eta_\ell}{\beta_\a}\gamma_\a\delta(t-t') \nonumber \\
& - s_\a(t)\sum_\ell \frac{\eta_\ell}{\beta_\a}s_\a(t')(\mu_\a-\im\zeta_\ell/\beta_\a)\ex^{-\im\phi_\a(t,t')}\ex^{-\im(\mu_\a-\im\zeta_\ell/\beta_\a)(t-t')}\gamma_\a .
\end{align}

The retarded component is also needed:
\begin{equation}
{\Sigma}_{\a,mn}^{1,\mathrm{R}}(t,t') = \sum_k E_{k\a}T_{mk\a}s_\a(t)g_{k\a}^{\mathrm{R}}(t,t')T_{k\a n}s_\a(t'),
\end{equation}
where the free Green's function takes the standard form
\begin{equation}
g_{k\a}^{\mathrm{R}}(t,t') = -\im\theta(t-t')\ex^{-\im\int_{t'}^t\ud\tb[ E_{k\a}+V_\a(\tb)]} = -\im\theta(t-t')\ex^{-\im\phi_\a(t,t')}\ex^{-\im E_{k\a}(t-t')}.
\end{equation}
Applying a similar procedure as above leads to
\begin{align}
{\Sigma}_{\a,mn}^{1,\mathrm{R}}(t,t') & = -\im s_\a(t) s_\a(t') \ex^{-\im\phi_\a(t,t')}\int_{-\infty}^\infty\frac{\ud\w}{2\pi}\w\ex^{-\im\w(t-t')}\underbrace{2\pi\sum_k T_{mk\a}\delta(\w-E_{k\a})T_{k\a n}}_{\approx\gamma_{\a,mn}}\theta(t-t') \nonumber \\
& = -\frac{1}{2} s_\a(t)s_\a(t')\ex^{-\im\phi_\a(t,t')}\frac{\partial}{\partial t'} \delta(t-t')\gamma_{\a,mn} .
\end{align}

In contrast to the electric-current formula~\cite{tuovinen_time-linear_2023-a}, the time-convolutions now involve derivatives of the Dirac delta function, $\frac{\partial}{\partial t'} \delta(t-t')$. The notation is purely formal and should be understood in the distributional sense: for two well-behaved functions, integration by parts gives $\int_a^b f'(x)g(x) \ud x = \left.f(x)g(x)\right|_a^b - \int_a^b f(x)g'(x)\ud x$. Here, these are Green's functions and self-energies [main text Eq.~(9)], and the convolution integrals have limits at $\pm \infty$. Since the functions vanish sufficiently fast at infinity, we obtain $\int_{-\infty}^\infty f'(x)g(x) \ud x = - \int_{-\infty}^\infty f(x)g'(x)\ud x$. For a distribution $g$ acting on $f$: $g[f] \equiv \int_{-\infty}^\infty f(x)g(x) \ud x$, the derivative of the distribution becomes $g'[f] \equiv \int_{-\infty}^\infty f(x)g'(x) \ud x = -\int_{-\infty}^\infty f'(x)g(x) \ud x = -g[f']$. Applying this to the delta function gives $\delta[f] = \int_{-\infty}^\infty f(x)\delta(x-x_0)\ud x = f(x_0)$ and $\delta'[f] = -\delta[f'] = -f'(x_0)$.

The time-convolutions involving derivatives of the Dirac delta function are then performed accordingly:
\begin{align}
& \int \ud \tb \ {\Sigma}_{\a}^{1,<}(t,\tb) G^{\mathrm{A}}(\tb,t) \nonumber \\
& = \int \ud \tb \frac{1}{2}s_\a(t)s_\a(\tb)\ex^{-\im\phi_\a(t,\tb)}\frac{\partial}{\partial \tb} \delta(t-\tb)\gamma_a G^{\mathrm{A}}(\tb,t) -\int\ud\tb 2\im s_\a(t)s_\a(\tb)\ex^{-\im\phi_\a(t,\tb)}\sum_\ell \frac{\eta_\ell}{\beta_\a}\gamma_\a\delta(t-\tb)G^{\mathrm{A}}(\tb,t) \nonumber \\
& - \int \ud \tb \ s_\a(t)\sum_\ell \frac{\eta_\ell}{\beta_\a}s_\a(\tb)(\mu_\a-\im\zeta_\ell/\beta_\a)\ex^{-\im\phi_\a(t,\tb)}\ex^{-\im(\mu_\a-\im\zeta_\ell/\beta_\a)(t-\tb)}\gamma_\a G^{\mathrm{A}}(\tb,t) \nonumber \\
& = - \frac{1}{2}s_\a(t)\gamma_\a\left.\frac{\partial}{\partial \tb}\left[s_\a(\tb)\ex^{-\im\phi_\a(t,\tb)} G^{\mathrm{A}}(\tb,t)\right]\right|_{\tb\to t} + s_\a^2(t)\sum_\ell \frac{\eta_\ell}{\beta_\a}\gamma_\a \nonumber \\
& - \sum_\ell\frac{\eta_\ell}{\beta_\a}(\mu_\a-\im\zeta_\ell/\beta_\a)\gamma_\a s_\a(t)\underbrace{\int \ud \tb \ s_\a(\tb)\ex^{-\im\phi_\a(t,\tb)}\ex^{-\im\mu_\a(t-\tb)}\ex^{-\zeta_\ell(t-\tb)/\beta_\a} G^{\mathrm{A}}(\tb,t)}_{\equiv \mathcal{G}_{\ell\a}(t)} ,
\end{align}
where we identified the embedding correlator $\mathcal{G}$ from Ref.~\cite{tuovinen_time-linear_2023-a}. Expanding the derivative in the above expression gives
\begin{align}
& \int \ud \tb \ {\Sigma}_{\a}^{1,<}(t,\tb) G^{\mathrm{A}}(\tb,t) \nonumber \\
& = - \frac{1}{2}s_\a(t)\gamma_\a \left. \left[\frac{\ud}{\ud\tb}s_\a(\tb)\right]\ex^{-\im\phi_\a(t,\tb)}G^{\text{A}}(\tb,t)\right|_{\tb\to t} - \frac{\im}{2}s_\a(t)\gamma_\a\left. s_\a(\tb)\left[-\im\frac{\partial}{\partial\tb}\ex^{-\im\phi_\a(t,\tb)}\right]G^{\text{A}}(\tb,t)\right|_{\tb\to t} \nonumber \\
& + \frac{\im}{2}s_\a(t)\gamma_\a\left. s_\a(\tb)\ex^{-\im\phi_\a(t,\tb)}\left[\im\frac{\partial}{\partial\tb}G^{\text{A}}(\tb,t)\right]\right|_{\tb\to t} + s_\a^2(t)\sum_\ell \frac{\eta_\ell}{\beta_\a}\gamma_\a - \sum_\ell\frac{\eta_\ell}{\beta_\a}(\mu_\a-\im\zeta_\ell/\beta_\a)\gamma_\a s_\a(t) \mathcal{G}_{\ell\a}(t) \nonumber \\
& = - \frac{\im}{4}s_\a(t)\gamma_a \dot{s}_\a(t) - \frac{\im}{2}s_\a(t)\gamma_\a \left. s_\a(\tb) V_\a(\tb)\ex^{-\im\phi_\a(t,\tb)} G^{\text{A}}(\tb,t)\right|_{\tb\to t} \nonumber \\
& + \frac{\im}{2}s_\a(t)\gamma_\a \left. s_\a(\tb)\left[\heff(\tb)G^{\text{A}}(\tb,t)+\delta(\tb-t)\right]\right|_{\tb\to t} + s_\a^2(t)\sum_\ell \frac{\eta_\ell}{\beta_\a}\gamma_\a - \sum_\ell\frac{\eta_\ell}{\beta_\a}(\mu_\a-\im\zeta_\ell/\beta_\a)\gamma_\a s_\a(t) \mathcal{G}_{\ell\a}(t) \nonumber \\
& = - \frac{\im}{4}s_\a(t)\gamma_a \dot{s}_\a(t) + \frac{1}{4}s_\a^2(t)\gamma_\a V_\a(t) - \frac{1}{4}s_\a^2(t)\gamma_\a \heff(t) + \frac{\im}{2}s_\a^2(t)\gamma_a\delta(0) \nonumber \\
& + s_\a^2(t)\sum_\ell \frac{\eta_\ell}{\beta_\a}\gamma_\a - \sum_\ell\frac{\eta_\ell}{\beta_\a}(\mu_\a-\im\zeta_\ell/\beta_\a)\gamma_\a s_\a(t) \mathcal{G}_{\ell\a}(t) .
\end{align}
The first and fourth terms in this result come with an explicit prefactor $\im$ with the remaining terms being purely real, so they will vanish when taking the real part for the energy current [cf.~Eq.~(8) in the main text].

The other time-convolution involving the retarded and lesser components can be calculated similarly:
\begin{align}
& \int \ud \tb \ {\Sigma}_{\a}^{1,\text{R}}(t,\tb) G^<(\tb,t) \nonumber \\
& = -\int \ud \tb \ \frac{1}{2} s_\a(t)s_\a(\tb)\ex^{-\im\phi_\a(t,\tb)}\frac{\partial}{\partial \tb} \delta(t-\tb)\gamma_\a G^<(\tb,t) = \frac{1}{2}s_\a(t)\gamma_a\left. \frac{\partial}{\partial \tb}\left[ s_\a(\tb)\ex^{-\im\phi_\a(t,\tb)}G^<(\tb,t)\right]\right|_{\tb\to t} \nonumber \\
& = \frac{\im}{2}s_\a(t)\gamma_a\left. \left[\frac{\ud}{\ud \tb}s_\a(\tb)\right]\ex^{-\im\phi_\a(t,\tb)}\left[-\im G^<(\tb,t)\right]\right|_{\tb\to t} - \frac{1}{2}s_\a(t)\gamma_a \left. s_\a(\tb)\left[-\im\frac{\partial}{\partial \tb}\ex^{-\im\int_{\tb}^t \ud\tau V_\a(\tau)}\right]\left[-\im G^<(\tb,t)\right]\right|_{\tb\to t} \nonumber \\
& - \frac{\im}{2}s_\a(t)\gamma_a \left. s_\a(\tb)\ex^{-\im\phi_\a(t,\tb)}\left[\im \frac{\partial}{\partial \tb} G^<(\tb,t)\right]\right|_{\tb\to t} \nonumber \\
& = \frac{\im}{2}s_\a(t)\gamma_a\dot{s}_\a(t) \rho(t) - \frac{1}{2}s_\a(t)\gamma_a \left. s_\a(\tb)\left[V_\a(\tb)\ex^{-\im\phi_\a(t,\tb)}\right]\left[-\im G^<(\tb,t)\right]\right|_{\tb\to t} \nonumber \\
& - \frac{\im}{2}s_\a(t)\gamma_a\left. s_\a(\tb)\ex^{-\im\phi_\a(t,\tb)}\left[\heff(\tb) G^<(\tb,t)\right]\right|_{\tb\to t} \nonumber \\
& = \frac{\im}{2}s_\a(t)\gamma_a\dot{s}_\a(t)\rho(t) - \frac{1}{2}s_\a^2(t)\gamma_a V_\a(t)\rho(t) + \frac{1}{2}s_\a^2(t)\gamma_a \heff(t) \rho(t) .
\end{align}
Here as well, the first term comes with an overall prefactor $\im$ with the remaining terms being purely real, so it will vanish when taking the real part for the energy current. Combining altogether gives for the energy current
\begin{align}\label{eq:energy-current-pole}
J_\a^{\text{E}}(t) & = 2\Re \mathrm{Tr}\left\{ - \frac{\im}{4}s_\a(t)\gamma_a \dot{s}_\a(t) + \frac{1}{4}s_\a^2(t)\gamma_\a V_\a(t) - \frac{1}{4}s_\a^2(t)\gamma_\a \heff(t) + \frac{\im}{2}s_\a^2(t)\gamma_a\delta(0) \right.\nonumber \\
& \left. + s_\a^2(t)\sum_\ell \frac{\eta_\ell}{\beta_\a}\gamma_\a - \sum_\ell\frac{\eta_\ell}{\beta_\a}(\mu_\a-\im\zeta_\ell/\beta_\a)\gamma_\a s_\a(t) \mathcal{G}_{\ell\a}(t) + \frac{\im}{2}s_\a(t)\gamma_a\dot{s}_\a(t)\rho(t) \right.\nonumber \\
& \left. - \frac{1}{2}s_\a^2(t)\gamma_a V_\a(t)\rho(t) + \frac{1}{2}s_\a^2(t)\gamma_a \heff(t) \rho(t) \right\} \nonumber \\
& = 2s_\a(t)\Re \mathrm{Tr}\left\{\gamma_a\left[s_\a(t)\frac{[\heff(t)-V_\a(t)\unity][2\rho(t)-\unity]}{4}-\sum_\ell\frac{\eta_\ell}{\beta_\a}\left((\mu_\a-\im\zeta_\ell/\beta_\a)\mathcal{G}_{\ell\a}(t)-s_\a(t)\right)\right]\right\}.
\end{align}

Eq.~\eqref{eq:energy-current-pole} is an explicit expressions for the energy current at the WBLA level. It is specified in terms of the one-electron density matrix $\rho$ and the embedding correlator $\mathcal{G}$~\cite{tuovinen_time-linear_2023-a}. In the main text, the generalization for finite bandwidth leads has been outlined, including more auxiliary correlators to be co-propagated, which reduces to this result at the proper limit (see Fig.~2 of the main text). All levels of the theory are implemented within the \textsc{cheers} nonequilibrium Green's function code~\cite{pavlyukh_cheers_2024-a}.

\section{Extracting Seebeck voltages}

\begin{figure}[h]
\center
\includegraphics[width=0.46\textwidth]{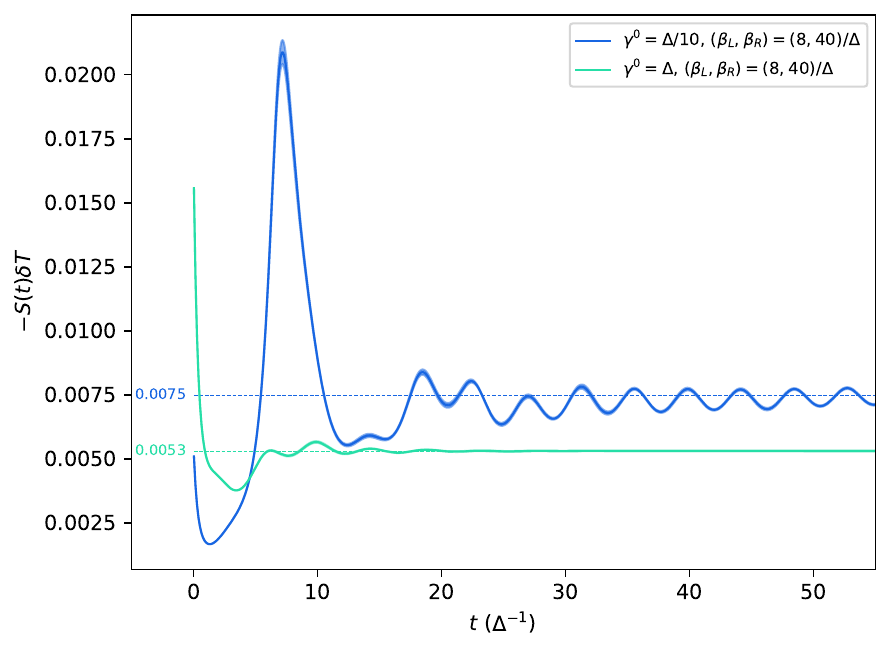}
\begin{caption}{
Extracted time-resolved Seebeck voltages (solid) and their mean values during the time evolution (dashed). The total time evolution continues outside the figure frame up to $t=122\Delta^{-1}=500$~a.u.
\label{fig:seebeckvoltage}}
\end{caption}
\end{figure}

The scheme for extracting the time-resolved Seebeck coefficient is explained with Fig.~8 in the main text. The corresponding voltage values at which the net charge current vanishes are required for the range of the voltage scans reported in Fig.~9 of the main text. For the two representative cases $(\beta_L, \beta_R) = (8, 40)/\Delta$ and $\gamma^0 = \{\Delta / 10, \Delta\}$, we show in Fig.~\ref{fig:seebeckvoltage} the extracted time-dependent voltages and the mean-values used for the voltage scan. The values in Fig.~\ref{fig:seebeckvoltage} correspond to the total voltage drop $\delta V=V_L^0 - V_R^0$, whereas the values reported in the inset of Fig.~8(a) in the main text correspond to $V\equiv V_L^0=-V_R^0$.

\section{Thermoelectric figure of merit}

We start from time-dependent charge and heat currents calculated under various voltage biases $\delta V$ and temperature gradients $\delta T$. In the linear-response regime, the currents can be expanded as
\begin{align}
J_{\mathrm{tot}}(t,\delta V,\delta T) &= a(t)\delta V + b(t)\delta T, \label{eq:linear1}\\
J^H_{\mathrm{tot}}(t,\delta V,\delta T) &= c(t)\delta V + d(t)\delta T,\label{eq:linear2}
\end{align}
where the coefficients $a,b,c,d$ can be obtained by a two-parameter linear regression at each time $t$. These coefficients represent the instantaneous Onsager's kinetic coefficients~\cite{callen_application_1948-a}:
\begin{equation}
a = \left.\frac{\partial J_{\mathrm{tot}}}{\partial \delta V}\right|_{\delta T=0}, \quad
b = \left.\frac{\partial J_{\mathrm{tot}}}{\partial \delta T}\right|_{\delta V=0}; \quad
c = \left.\frac{\partial J^H_{\mathrm{tot}}}{\partial \delta V}\right|_{\delta T=0}; \quad
d = \left.\frac{\partial J^H_{\mathrm{tot}}}{\partial \delta T}\right|_{\delta V=0}.
\end{equation}
Since they are instantaneous, the procedure can be performed for each time step, and thereby we can define the time-dependent transport coefficients under standard conditions.
Seebeck coefficient (open circuit, $J_{\mathrm{tot}}=0$, as done earlier):
$S(t) = -\left.\frac{\delta V}{\delta T}\right|_{J_{\mathrm{tot}}=0} = \frac{b(t)}{a(t)}$.
Electrical conductance (isothermal, $\delta T=0$):
$\mathcal{G}(t) = a(t)$.
Thermal conductance (open circuit, $J_{\mathrm{tot}}=0$):
Using $J_{\mathrm{tot}}=0 \Rightarrow \delta V=-\frac{b}{a}\delta T$, we get
$\mathcal{K}(t) = \left.\frac{J^H_{\mathrm{tot}}}{\delta T}\right|_{J_{\mathrm{tot}}=0} = d(t)-\frac{c(t)b(t)}{a(t)}$.
Finally, the time-resolved thermoelectric figure of merit is
\begin{equation}
ZT(t) = \frac{\mathcal{G}(t)\,S(t)^2\,T}{\mathcal{K}(t)}
       = \frac{b(t)^2 T}{a(t) d(t)-b(t) c(t)},
\end{equation}
where the temperature $T$ is taken as the average value within each linear regression. The error bars of this procedure can be obtained via the propagation of uncertainty:
\begin{equation}
\delta (ZT) = \sqrt{\left(\frac{\partial (ZT)}{\partial a}\delta a\right)^2+\left(\frac{\partial (ZT)}{\partial b}\delta b\right)^2+\left(\frac{\partial (ZT)}{\partial c}\delta c\right)^2+\left(\frac{\partial (ZT)}{\partial d}\delta d\right)^2},
\end{equation}
where $\delta a,\delta b,\delta c,\delta d$ are the uncertainties of each coefficient extracted from Eqs.~\eqref{eq:linear1} and~\eqref{eq:linear2}.

\begin{figure}[h]
\center
\includegraphics[width=0.46\textwidth]{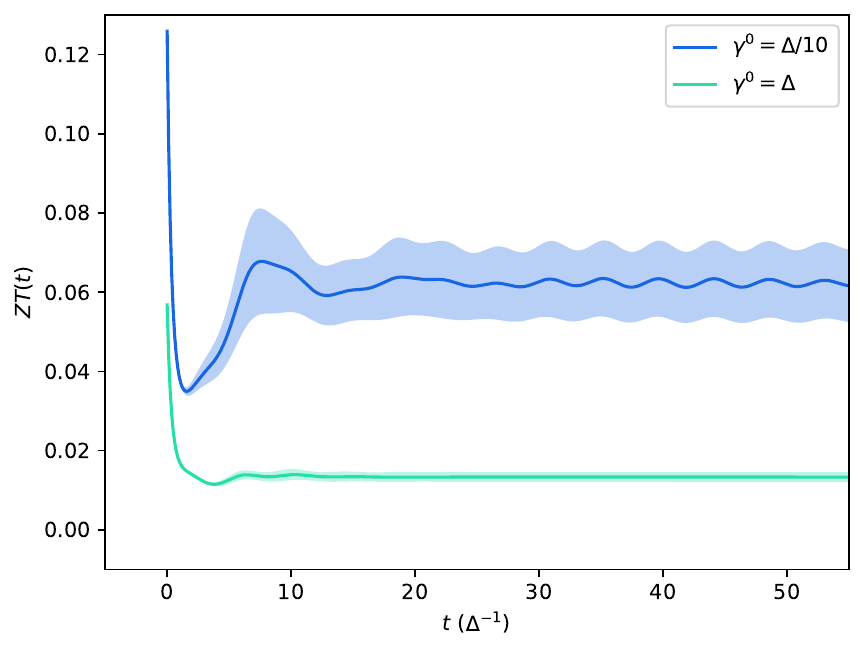}
\begin{caption}{
Time-dependent thermoelectric figure of merit obtained by linear regression using Eqs.~\eqref{eq:linear1} and~\eqref{eq:linear2}. The underlying time-dependent charge and heat current data are the ones regarding Fig.~8 of the main text. The shaded areas represent uncertainty intervals of the linear regression.
\label{fig:zt}}
\end{caption}
\end{figure}

For the cyclobutadiene molecular junction, we show in Fig.~\ref{fig:zt}, the time-dependent thermoelectric figure of merit for the time-dependent charge and heat current data regarding Fig.~8 of the main text. The extracted figure of merit $ZT(t)$ falls in the typical range for molecular junctions, $ZT \sim 0.01$--$0.1$, reflecting the generally modest thermoelectric performance of single-molecule devices due, e.g., to contact resistance effects~\cite{wang_thermal_2020-a}. We find that the figure of merit increases as the tunneling rate between the molecule and the leads is reduced, consistent with the increased thermoelectric energy conversion efficiency in the weak-coupling regime. Furthermore, the time-dependent profile of $ZT(t)$ is roughly similar to that of the Seebeck coefficient $S(t)$, cf.~Fig.~8 in the main text, since both are governed by the same open-circuit condition. However, the error bars of $ZT$ are slightly larger because there are only two different values for the temperature gradients in the data set, cf.~Fig.~8 in the main text.

Following Ref.~\cite{benenti_thermodynamic_2011-a}, the maximal thermoelectric efficiency in the linear-response regime is $
\eta_{\max} = \eta_C \frac{\sqrt{1+ZT}-1}{\sqrt{1+ZT}+1}$. For small $ZT$ this reduces to $\eta_{\max} \approx \eta_C \frac{ZT}{4}$. Using the lead temperatures $(\beta_L, \beta_R) = (8, 40)/\Delta$, the Carnot efficiency evaluates to $\eta_C=0.8$. Thus, in our setup, this approach would predict maximal efficiencies $\eta_{\max} \sim 0.2 ZT$. Comparing with Fig.~9 in the main text, the estimated values for the thermoelectric energy conversion efficiency are in the same range. However, these two alternative approaches answer related but distinct questions~\cite{muralidharan_performance_2012-a}. The $ZT$-based approach provides an idealized, linear-response upper bound, assuming small temperature and voltage differences. In this framework, weak coupling sharpens the molecular transmission, enhances the Seebeck coefficient, and increases $ZT$, leading to higher maximal efficiency. In contrast, the circuit-based, operational efficiency, cf.~Fig.~9 in the main text, utilizes the actual power extracted from the junction under a finite load. In this situation, strong coupling increases the current through the junction, which can outweigh the smaller Seebeck coefficient, yielding a higher practical efficiency. As a result, the two methods can therefore differ to some extent: $ZT$ predicts efficiency based on intrinsic molecular properties, while the operational approach reflects realistic power extraction in a measurable circuit configuration.


%

\end{document}